\documentclass{IEEEcsmag}

\usepackage[colorlinks,urlcolor=blue,linkcolor=blue,citecolor=blue]{hyperref}

\usepackage{upmath}
\usepackage{float}
\usepackage{tcolorbox}
\usepackage{xcolor}
\usepackage{booktabs}
\usepackage{multirow}
\usepackage{array}
\usepackage{tabularx}
\usepackage{fontawesome5}
\usepackage{bibunits}

\jvol{XX}
\jnum{XX}
\paper{8}
\jmonth{}
\jname{IT Professional}
\pubyear{2025}

\definecolor{themeorange}{rgb}{0.98,0.64,0.098}

\begin{document}

\sptitle{THEME: Inclusive data experiences}

\title{PLUTO: A Public Value Assessment Tool}

\author{\text{L}aura Koesten}
\affil{University of Vienna, Vienna, 1090, Austria}

\author{P{\'e}ter Ferenc Gyarmati}
\affil{University of Vienna, Vienna, 1090, Austria}

\author{Connor Hogan}
\affil{University of Vienna, Vienna, 1090, Austria}

\author{Bernhard Jordan}
\affil{University of Vienna, Vienna, 1090, Austria}

\author{Seliem El-Sayed }
\affil{University of Vienna, Vienna, 1090, Austria}

\author{Barbara Prainsack}
\affil{University of Vienna, Vienna, 1090, Austria}

\author{Torsten M{\"o}ller}
\affil{University of Vienna, Vienna, 1090, Austria}

\markboth{THEME}{Inclusive data experiences}

\begin{abstract}
We present PLUTO (\textbf{P}ublic Va\textbf{LU}e Assessment \textbf{TO}ol), a framework for assessing the public value of specific instances of data use. Grounded in the concept of data solidarity, PLUTO aims to empower diverse stakeholders--including regulatory bodies, private enterprises, NGOs, and individuals--to critically engage with data projects through a structured assessment of the risks and benefits of data use, and by encouraging critical reflection.

This paper discusses the theoretical foundation, development process, and initial user experiences with PLUTO. Key challenges include translating qualitative assessments of benefits and risks into actionable quantitative metrics while maintaining inclusivity and transparency. Initial feedback highlights PLUTO's potential to foster responsible decision-making and shared accountability in data practices.
\end{abstract}

\maketitle

\section{Introduction}

Data has become a fundamental part of modern life, shaping how we work, live, and connect with the world. However, access to the benefits of data use and exposure to its risks remains deeply inequitable. General (non-technical) audiences often face the most significant risks, 
while powerful actors disproportionately benefit. Moreover, less technical communities are frequently excluded from discussions on data governance and lack access to remedies when harm occurs \cite{mcmahon2020big}. These exclusions fall especially heavily on marginalized groups, who face the sharpest consequences of data-related harms, such as algorithmic discrimination or privacy violations.

At the core of these disparities lies a shared challenge: organizations, regulators, and the public alike struggle to determine whether specific uses of data genuinely serve society. Assessing the risks and benefits of data use for public value is difficult in practice, and current regulatory efforts often rely on the vague and binary notion of public interest. 

Addressing this gap requires data governance centered on public purpose rather than private profit \cite{kickbusch2021lancet}. Solidarity-based data governance provides such a foundation by seeking to ensure that the benefits and risks of digital practices are shared collectively and fairly, emphasizing public value in the process \cite{prainsack2022white,Hogan2025}. 
In the data solidarity framework, public value arises when data use can plausibly be assumed to benefit the public without causing significant undue harm to any person or group~\cite{prainsack2022white}. Value is maximized when benefits are inclusive, reaching diverse groups across society. Building on this  foundation, the central problem this work addresses is how to meaningfully operationalize and assess the public value of data use. 

To this end, we introduce the Public Value Assessment Tool (PLUTO), a structured questionnaire designed to make the concept of public value actionable by evaluating the public value that a specific data use is likely to generate~\footnote{\url{https://pluto.univie.ac.at}}.

The tool enables individuals and organizations to conduct an assessment of the public value of specific data uses (e.g., in a particular project or organization) through a guided questionnaire that invites reflection and evaluation. Expert knowledge of data analytics or governance is not required. 
Representatives of companies or public authorities can use it to evaluate and understand how their data practices could foster greater public value, for instance, while citizens can use it to help decide whether to share their data with companies or research organizations.

This paper discusses the background, development, and initial user experiences with PLUTO through the lens of user-centered design approach to benefit both survey developers and users.

Our contributions are along three lines:

\noindent\textbf{The design and development of PLUTO:} We present the theoretical foundations, conceptualization, and iterative development of PLUTO (\textbf{P}ublic Va\textbf{LU}e Assessment \textbf{TO}ol). 
PLUTO employs a structured algorithmic and question-based framework accessible to diverse stakeholders, including those without advanced technical expertise. The tool integrates a questionnaire with a tailored visualization component for real-time feedback, which was evaluated in several rounds, including expert feedback and a usability study. It was developed using principles of inclusivity and adaptability to effectively address the diverse needs of stakeholders across global contexts.

\noindent\textbf{Reflections on translating qualitative assessments into quantitative metrics:} We discuss the challenges inherent in quantifying complex, multi-faceted concepts such as the benefits, risks, and public value of data use. 
Our reflections offer a foundation for bridging critical data studies with practical, scalable tools for data governance.

\noindent\textbf{Insights from real-world feedback on PLUTO:} Through iterative feedback collected from diverse experts and multi-disciplinary stakeholders in policy, data science, law, and ethics, we emphasize the critical importance of inclusivity in tool design. This feedback informed iterations of PLUTO, ensuring that the tool acknowledges user diversity in regional, technical, and cultural contexts. 
Updates improved language accessibility, user experience, bias reduction, and transparency through features like information boxes and weight rationales. These changes highlight the need for collaborative stakeholder engagement to build trust, ensure contextual fairness, and foster meaningful use of tools like PLUTO.

The following section provides context on the notion of data solidarity, understanding public value within this framework, and the initial conceptualization and development of the tool. While another paper \cite{El-Sayed31122025} introduced the conceptual considerations and commitments that underpin the development of the tool, in this paper, we detail the user-centered design, iterative visualization development, technical implementation, and empirical usability evaluation of PLUTO as an inclusive data experience. Next, the methods section outlines the design process of the tool itself, with a focus on the visualization aspect of the tool. We then discuss the initial evaluation and refinement of PLUTO, highlighting efforts to ensure the tool's relevance for less technical and more marginalized communities and users with diverse knowledge backgrounds. Finally, we reflect on PLUTO's reception and discuss future directions.

Through the development of PLUTO, we show how a structured questionnaire can empower diverse audiences, including non-experts, to engage meaningfully in data governance.

\section{Background}
\label{sec:background}

\vspace{1em}
Data solidarity is an approach to data governance that seeks to achieve a more equitable sharing of benefits and harms emerging from digital practices \cite{prainsack2022white}. 

It rests on three pillars (see \cite{El-Sayed31122025}): promoting data use with public value, mitigating harm, and sharing commercial profits with communities.


A key premise of data solidarity is that it does not assume that risk lies in data \textit{types}, but in data uses: how data is used, by whom, and for whose benefit. Data solidarity distinguishes between four types of data use:

\noindent\textbf{Type A}: Likely creates significant public value as it will plausibly yield significant benefits without posing unacceptably high risks. These types should be supported (Pillar I)

\noindent\textbf{Type B}: Unlikely to yield significant public benefits, but also poses minimal risks. Financial profits from Type B uses should be partially returned to the public domain (Pillar III)

\noindent\textbf{Type C}: Produces significant public benefits, but poses unacceptably high risks. Type C uses are only permissible if risks can be reduced to acceptable levels (Pillar II)

\noindent\textbf{Type D}: Likely does not create significant public value while at the same time posing unacceptably high risks. These activities should be banned (Pillar II)

\noindent Across all types of data use, harm mitigation measures should be in place to support those who experience negative outcomes due to data use (Pillar II) \cite{mcmahon2020big}.

\begin{figure}
    \centering
    \includegraphics[width=1\columnwidth]{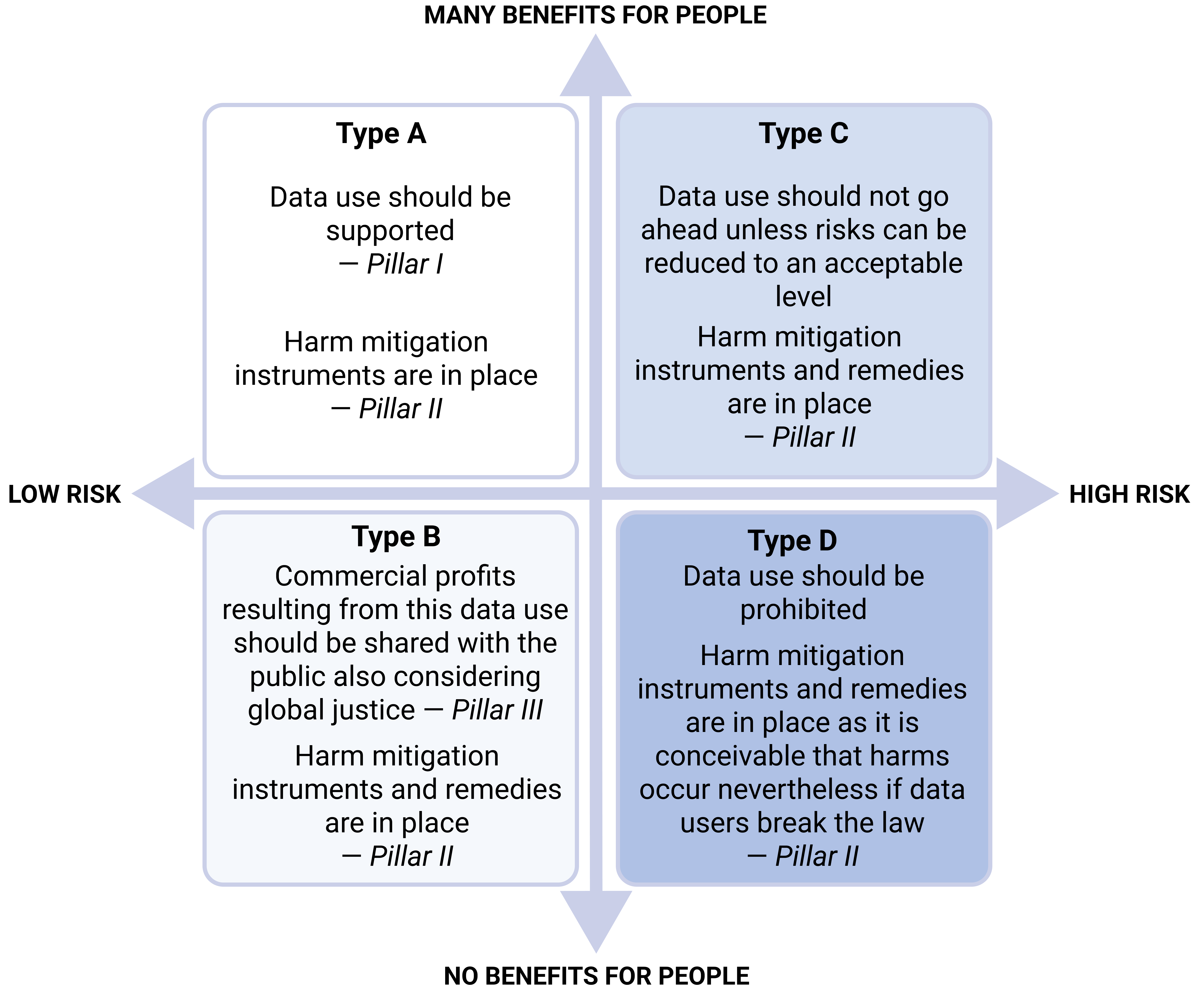}
    \vspace{-0.75em}
    \caption{The four types of data use within data solidarity framework, plotted on a risk-benefit matrix. Each quadrant defines a data use type (A-D) and prescribes a corresponding governance action—from support (Type A) to prohibition (Type D)—based on its position within the framework.}
    \label{fig:data_solidarity_matrix}
\end{figure}

We developed a four-quadrant scheme to visually depict the different types of data use within data solidarity (see \autoref{fig:data_solidarity_matrix}) \cite{prainsack2022white}. This would influence the final results page of PLUTO (discussed below), as it offers a clear, practical visualization of the framework underlying public value assessment.

Data solidarity aims to promote data use with high public value while clearly and effectively prohibiting activities that present unacceptable risks to individuals or society at large.

\subsection{Public value}

Within data solidarity, public value is realized when \cite{Hogan2025}: (1)~it contributes towards widely agreed upon goals for society; (2)~no person or group will likely risk significant and undue harm; (3)~there are sufficient harm mitigation procedures in place; (4)~it is done through a publicly transparent process; and (5)~it is primarily motivated by an intention to benefit society or a subsection of society.

Inclusivity is vital for underserved or marginalized communities, where limited resources mean even modest benefits can have a transformative impact \cite{prainsack2016thinking}. Public value  increases when the benefits are distributed fairly (i.e., by helping bridge the power gap between individuals and their communities on the one hand and large digital entities on the other, and/or by helping reduce regional or global inequality), and when marginalized groups are likely to experience them. It is further strengthened when benefits are sustainable (in that its value will foreseeably extend into the future or to future generations).


We chose to focus on "public value" as opposed to the conventional notion of "public interest", which predominates previous approaches to data governance. The concept of public interest has been critiqued as "inconsistent and ambiguous" \cite{fukumoto2019public}, with similar critiques directed at related notions such as the "common good" \cite{bryson2014public}. Employing a conception of public value allows for a more nuanced evaluation of data uses, circumventing the need for an exhaustive and generic list of data applications deemed in (or not in) the public interest.

\subsection{PLUTO: Concept and initial development}
The motivation for PLUTO stemmed from the need for a practical means to assess the public value of different instances of data use to advance the goals of data solidarity.
Starting from the need to assess data uses instead of data types, we based our approach on existing impact assessments for digital applications.
 
Inspired by the structured format of Algorithmic Impact Assessments, we developed PLUTO as a freely accessible online questionnaire.

\begin{bibunit}[IEEEtran]
\begin{tcolorbox}[
    colback=themeorange!15,
    colframe=themeorange,
    title=\textbf{Inspiration: Algorithmic Impact Assessments},
    fonttitle=\bfseries,
    sharp corners,
    boxsep=3pt,
    left=3pt,
    right=3pt,
    top=3pt,
    bottom=3pt
]
Algorithmic Impact Assessments (AIAs) are frameworks for evaluating the societal harms of automated systems by considering their effects on fairness and transparency \cite{reisman2018algorithmic, metcalf2021algorithmic}. Pioneering government-led examples from Canada and New Zealand use structured questionnaires to score risk and guide responsible deployment \cite{metcalf2021algorithmic}.

\vspace{2mm}
{\footnotesize
\setlength{\parskip}{2pt}
\def\section*#1{}
\putbib[references-sidebar]}

\end{tcolorbox}
\end{bibunit}

Once the questionnaire structure was established, we drafted and reviewed a preliminary set of questions internally. Guided by our operationalization of public value and the pillars of data solidarity \cite{El-Sayed31122025}, we sought in part to use the weighted-question format to reward data uses in which the benefits were inclusive, reaching marginalized and underrepresented communities, and those data uses in which measures were taken to reduce risks for vulnerable populations. We also aimed to promote data use to address global challenges, such as climate change, for their broader societal contributions.


In assessing the benefits and risks, a key challenge was recognizing that impacts extend across different populations and time horizons. Harms are not limited to immediate, direct consequences for individuals, such as a privacy breach. Broader, systemic risks can emerge that affect entire communities, including those who did not provide data. For instance, biased algorithms can perpetuate discrimination, disproportionately harming marginalized groups. Importantly, significant harms can also arise from lawful data practices, which may exacerbate societal inequalities or enable surveillance, often without clear avenues for remedy for those affected.

Similarly, the benefits of data use can manifest across various timescales and levels of societal impact. Some benefits are immediate, such as actionable insights that improve public services. Others are long-term and collective, emerging from data-driven research into complex challenges like climate change or public health, with outcomes that may only be realized in the future.


To account for this complexity, the questionnaire was designed to adopt a far-reaching approach when evaluating benefits and risks. We asked users to consider the plausible benefits of their data use for the environment, marginalized communities, and future generations, as well as to whom these benefits mainly accrued (e.g., to the person or organization using the data, to users of a product/service emerging from the data use, or to non-users). Users 
-- who can be any natural or legal person using data -- would generally score higher if their answers indicated that public benefits were high and fairly distributed.

Regarding potential risks to people and communities, the first iteration of the questionnaire listed physical, financial, and informational risks as options to choose from (this was later expanded), with an additional "other" option through which users could make suggestions. Users were also asked to consider the plausible risks, broadly conceived, to the environment and marginalized communities of the data use and steps taken to assess, reduce, and communicate risks to the public. The data user's public value score would be lower, all things being equal, when the risks were more numerous and higher, and when marginalized communities and the environment were at an elevated risk of harm. Both the descriptions of benefits and risks remain fairly general and will need additions and refinement depending on the intended context of using PLUTO.

Furthermore, given the emphasis of data solidarity on reducing power asymmetries, it was necessary to include questions that assessed the relative size and power of the data user within the public sphere. This would require gathering information about the entity or person being assessed. To this end, we included questions on the primary activity of the data user (e.g., selling a product or service, conducting research), their financial structure (e.g., main sources of funding), and their public reporting requirements. Larger, more well-resourced users would generally have a higher public value burden than smaller users (i.e., the latter would score higher for these questions).

Finally, in line with Pillar II of data solidarity, we determined that questions related to harm mitigation measures (in addition to institutional safeguards that reduce \textit{risk}) should also be included. These included asking the user if harms were monitored, if their activities could be stopped immediately upon the discovery of harm, and if there was a complaints procedure in place for users who experience harm. Users' public value scores would be improved if robust, effective support measures were implemented for users who experience harm.
The questions were then refined through several rounds of internal feedback. Once the initial structure and content for the questionnaire were settled, the tool and accompanying website were constructed.

\vspace{5px}
\section{Methods}

Prior works in the VIS and HCI community provide evidence for the importance of visualizations in helping diverse users interpret complex information \cite{yen2020decipher} and recognize the challenge of creating accessible tools, evaluating complex interactive systems, and translating qualitative concepts into quantitative, visual representations \cite{crain2023topicvis}. Our design, development, and evaluation process was guided by these principles.

\subsection{Iterative Design Process}

The tool was developed through an iterative, user-centered design process. This involved in-depth discussions between the visualization and social science experts on the author team, combined with multiple rounds of feedback from domain experts and end-users. We began by co-designing low-fidelity prototypes to explore different ways of representing risk and benefit, which were then refined into a high-fidelity interactive tool based on structured feedback (see \hyperref[sec:evaluation]{Evaluation Section}). This process enabled the final design to be grounded in both its theoretical framework and the practical needs of its users.

\subsection{Summary of requirements}

The tool's requirements were informed by the primary objective of PLUTO: to provide a means of evaluating the public value of instances of data use in a quantitative and question-based manner, emphasizing inclusivity. PLUTO is intended for a diversity of stakeholders, including regulatory bodies, private enterprises, individuals, companies, and NGOs. We identified the requirements for an online questionnaire application consisting of two key components: a user-friendly survey interface that allows participants to submit their responses and a visualization component that enables participants to interpret their survey results and gain actionable insights. These insights should guide users in enhancing the benefits emerging from their data use while reducing the risks and mitigating associated harms.
The data to be visualized, the questionnaire score, is a composite value of score tuples (risk, benefit) across four dimensions listed in \autoref{fig:questionnaire_dimensions}:
\hfill
\begin{figure}[h!]
    \centering
    \begin{minipage}{0.6\textwidth}
        \centering
        \begin{enumerate}
            \item[i.)] Information About the Applicant
            \item[ii.)] Benefits of the Applicant's Activity
            \item[iii.)] Risks of the Applicant's Activity
            \item[iv.)] Institutional Safeguards
        \end{enumerate}
    \end{minipage}
    \caption{Key elements for assessing the public value of data use. These elements align with the categories of questionnaire items.}
    \label{fig:questionnaire_dimensions}
\end{figure}
\hfill

Our goal was to employ a visualization design that is easily comprehensible to an audience without requiring prior knowledge of data analysis or a quantitative background. We sought a visualization technique that is informative yet inclusive, ensuring accessibility and understanding across diverse user groups regardless of their data literacy.

\subsection{Previous work}

We reviewed existing literature and tools related to online questionnaire solutions and visualization techniques that enable users to analyze and understand the implications of their responses.


\subsubsection{Questionnaire Component} While several open-source tools provide robust solutions for survey creation, they primarily focus on enabling survey designers to analyze aggregated results. As PLUTO's core objective is to provide immediate, individualized feedback to the survey respondent, existing platforms were not fully suited to our needs. This required a hybrid approach: building a custom visualization layer on top of a standard survey engine to deliver tailored, actionable insights directly to the user.
We envisioned PLUTO as a unified tool where the formal questionnaire component and the interactive survey app are synchronized by design, facilitating future development and adaptation by other organizations. To achieve this, we prioritized programmatic questionnaire frameworks that offer high customizability. To support broad adoption and ensure data security, we integrated the permissively-licensed SurveyJS framework and designed the system for self-hosting.

\subsubsection{Visualization Component} Various visualization techniques exist for effectively representing multivariate data, including radar charts~\cite{Chambers1983}, which display multiple variables on a single chart with each variable represented by a spoke radiating from the center; stacked bar charts, which illustrate the composition of a total score; and heatmaps, which use color intensity to visualize scores across multiple components. However, we could not identify any existing solution that simultaneously satisfies all the above-summarized requirements.

\subsection{Prototypes}

In our exploration of visualization prototypes, we broadly considered three main approaches to effectively convey the results of submissions: single-value (1D), double-value (2D), and quadruple-value (4D) visualizations. Each approach is designed to capture and present the data's complexity differently while striving for clarity and interpretability. We outline the characteristics of these approaches to illustrate our reasoning and iterative design process.

\begin{figure*}
    \centering
    \includegraphics[width=1\textwidth]{figures/chart-prototypes/Prototypes.overview.5x5.highlighted.png}
    \caption{An overview of low-fidelity prototypes for visualizing submission results. The design highlighted in the orange frame was selected for implementation. All designs are available at \faGithub~\href{https://github.com/PLUTO-UniWien/PLUTO/blob/f300cb595e3e5618c9af8f6cdcef9c83a5730144/docs/designs/Prototypes.pdf}{PLUTO-UniWien/PLUTO/docs/designs}.}
    \label{fig:prototypes:overview}
\end{figure*}

\subsubsection{Single-Value Prototypes (1D)}

The 10 single-value prototypes (see the overview in \autoref{fig:prototypes:overview} and details in \autoref{fig:prototype:singlevalue1} -- \ref{fig:prototype:singlevalue10}) are designed to distill the submission results into a unified score, offering an overall view of the respondent's performance. This approach simplifies the data by projecting various components onto a single dimension. By calculating a score that reflects both the maximum reachable score and the actual score achieved, these visualizations provide an easy-to-understand summary that enables quick and intuitive interpretation. The design further incorporates detailed visual elements (as a detail-on-demand approach in the form of companion prototypes \cite{shneiderman2003eyes}) to enhance understanding of the score's composition, allowing users to grasp the nuances of their performance at a glance.

\subsubsection{Single-Value: Detail Companion Prototypes}

To address the potential oversimplification inherent in reducing scores to a single dimension, we developed detail companion prototypes for the Single-Value visualizations (see the overview in \autoref{fig:prototypes:overview} and details in \autoref{fig:prototype:detail1} -- \ref{fig:prototype:detail5}). These prototypes provide additional context and insight into the score's composition, enabling a more comprehensive understanding of the underlying data.

\subsubsection{Double-Value Prototypes (2D)}

Moving beyond a singular dimension, the double-value prototypes (see the overview in \autoref{fig:prototypes:overview} and details in \autoref{fig:prototype:doublevalues1} -- \ref{fig:prototype:doublevalues3}) introduce a dual-dimensional perspective that captures the interplay between "Benefits" and "Risks". This visualization strategy aims to highlight the relationship between these two key dimensions, offering insights into how they co-exist and influence each other. By mapping these aspects onto a two-dimensional plane, users can identify and interpret potential trade-offs, allowing for a more comprehensive analysis of the submission results.

\subsubsection{Quadruple-Value Prototypes (4D)}

The quadruple-value prototypes (see the overview in \autoref{fig:prototypes:overview} and details in \autoref{fig:prototype:quadruplevalues1} -- \ref{fig:prototype:quadruplevalues6}) take a multidimensional approach, representing relationships among the four distinct sections of the survey: "Information about the Applicant" (i.e., as noted above, the relative size and power of the data user relative to the public), "Benefits of the Applicant's Activity", "Risks of the Applicant's Activity", and "Institutional Safeguards". Each prototype variant seeks to visualize the complex interplay between these sections, thereby capturing the multifaceted nature of the data. Through these visualizations, we aim to provide a holistic view that highlights interdependencies and contrasts across different dimensions, aiding stakeholders in making informed decisions based on a comprehensive data assessment.

\subsubsection{Rationale}
After an initial round of internal feedback, we selected the  \textbf{Quadrants with Color Coding (Double Values 2)} (\autoref{fig:prototype:doublevalues2}) as the foundational component for our results visualization to balance the need for nuanced, multidimensional feedback with the core requirement of accessibility for a non-expert audience. First, the 2D quadrant leverages the high perceptual accuracy of judging position along common scales. Foundational work by Cleveland and McGill has shown this to be a more accurate elementary perceptual task than judging angle or area \cite{cleveland1984}. We therefore prioritized this positional encoding over alternatives like radial charts, which rely on less accurate angle and area judgments. This ensures that the core trade-off between risk and benefit is represented in a way that minimizes perceptual error. Second, the 2x2 matrix is a powerful and pervasive analytical idiom, which lowers the cognitive barrier for entry. This format is widely used to help users structure complex trade-offs in diverse contexts, from the Boston Consulting Group (BCG) Matrix\footnote{\url{https://bcg.com/about/our-history/growth-share-matrix}} in business strategy, to the Eisenhower Matrix\footnote{\url{https://eisenhower.me/eisenhower-matrix/}} in time management, and SWOT analysis\footnote{\url{https://investopedia.com/terms/s/swot.asp}} for strategic planning. This deep familiarity allows users to apply existing mental models to our tool, a finding supported by research showing that lay audiences comprehend conventional visualization formats more effectively than novel ones \cite{kennedy2016visconventions,lee2015people}. Third, we acknowledge that while a 2D plot excels at showing relationships, extracting precise numerical values from axes can be difficult for some users. Our final design mitigates this weakness by complementing the 2D quadrant with two separate 1D sliders that display the exact scores for risk and benefit (\autoref{fig:resultpage}). This composite design provides both a qualitative \textit{overview} (the quadrant) and quantitative \textit{details-on-demand} (the sliders), a core principle of effective information visualization \cite{shneiderman2003eyes}. This dual representation caters to different user needs simultaneously without requiring interaction. Finally, color is used deliberately as a categorical overlay to segment the space into the four PLUTO data use types, not to encode a precise value—a task for which color saturation is perceptually ill-suited \cite{cleveland1984}. This supports rapid, pre-attentive processing, allowing users to immediately grasp the result's classification. This synthesis of perceptual accuracy, a familiar matrix convention, and a composite overview-plus-detail view creates a visualization that is both nuanced and promotes comprehension among its intended non-specialist audience.

\subsection{Design and features of the final tool} 

The final tool was implemented as a web-based application.\footnote{\href{https://pluto.univie.ac.at}{https://pluto.univie.ac.at}} Before starting the questionnaire, users are presented with introductory information about the topic and the tool, a privacy policy, an imprint, citation guidelines, and a button to start the survey.

\begin{figure}
	\centering	
 \includegraphics[width=\columnwidth]{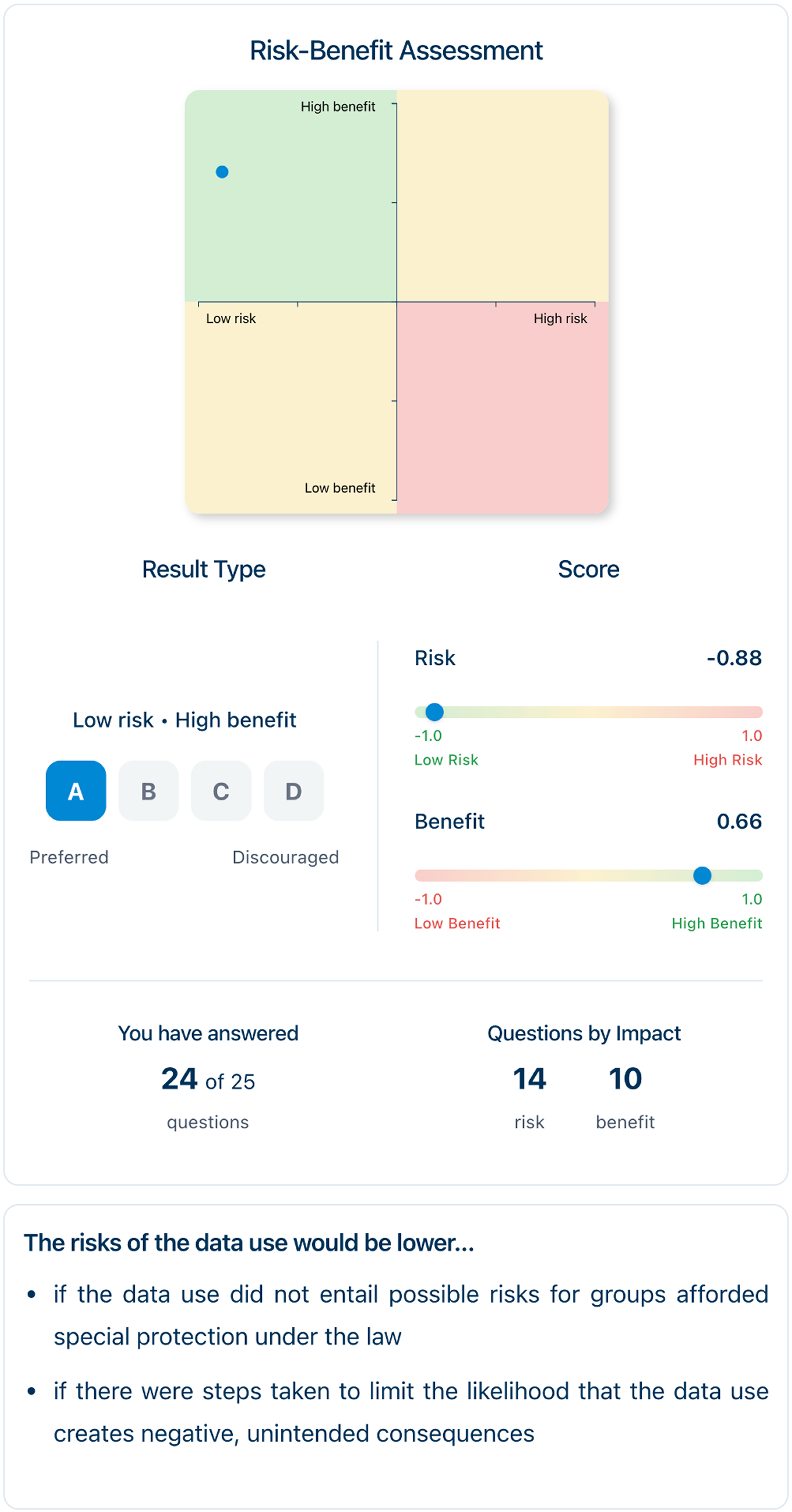}
	\caption{PLUTO results interface, integrating multiple components to support user comprehension. The primary visualization is a risk-benefit matrix that plots the overall assessment, contextualized by a qualitative A-D scale (from favorable to discouraged). To enhance accessibility for users less familiar with coordinate systems, separate sliders offer a direct, one-dimensional view of the Risk and Benefit scores. For full transparency, the interface details how user answers contribute to the final result. Finally, actionable recommendations provide concrete steps for improving the project's public value by increasing benefits and mitigating risks.
    }
	\label{fig:resultpage}
\end{figure}

The tool asks 25 questions across the four dimensions for assessing the public value of data use outlined by \autoref{fig:questionnaire_dimensions}.
The survey is designed to display each question on a separate page. Users can navigate to previous questions and access a preview page, which provides an overview of all questions and their selected answers. No evaluations or feedback are shown during the survey; results are presented only at the end after submission. After completing the PLUTO questionnaire, the tool generates a unique position in a multidimensional coordinate system (see \autoref{fig:resultpage} in the online supplement). The two axes represent the associated risks and benefits of the data use, respectively. The resulting position falls into one of four quadrants (A-D, see \autoref{fig:data_solidarity_matrix}), each linked to a specific data use category and a set of rules based on Data Solidarity Pillars I-III (see \autoref{tab:data_solidarity_pillars}). Users can export these results to PDF for sharing and further consideration. More detailed feedback is provided on the result page, explaining ways to improve the score to users. A contextual explanation of public value is included to help interpret the results and offer further insights. As shown in \autoref{fig:questionnaire_item}, if a respondent indicates a lack of procedures for responding to public information requests about data activities, the tool provides the direct recommendation: \textit{"The risks of the data use would be lower if the data user complied (better) with information requests regarding data use."}  This targeted approach helps users pinpoint areas needing attention and understand how specific actions can influence their public value score.

\begin{figure}[htbp]
    \centering
    \fbox{%
        \begin{minipage}{0.95\linewidth}
            \vspace{0.5em}
            \textbf{Q8.} Does the data user have a procedure in place to respond to public requests for information on its activities (e.g., freedom of information requests, other legal obligations, or voluntarily)?

            \vspace{1em}
            \textit{Answer Impact: $x$-axis}

            \vspace{1em}
            \begin{tabular}{p{0.75\linewidth}l}
                \toprule
                \textbf{Answer Choice} & \textbf{Weight} \\
                \midrule
                Procedure in place to respond to public requests for information & +2 \\
                No procedure in place to respond to public requests for information & -2 \\
                None of the above (please specify) & 0 \\
                I don't know & - \\
                \bottomrule
            \end{tabular}

            \vspace{1.5em}

            \textbf{Explanation:} Whether or not the data user responds to public requests for information matters when assessing public value. Having a procedure in place is more likely to lead to data use with a high public value due to increased public accountability (+2). Having no procedure is less likely to lead to low public value (-2).

            \vspace{1em}

            \textbf{Recommendation:} The risks of the data use would be lower if the data user complied (better) with information requests regarding data use.

            \vspace{0.5em}
    \end{minipage}%
    }
    \vspace{0.75em}
    \caption{Example of a questionnaire item with a contextual explanation of weighting and textual recommendation on how a better score could be achieved through more appropriate data practices. This example is drawn from Question 8 of the survey, impacting the risk score associated with the user's data use, visualized on the $x$-axis.}
    \label{fig:questionnaire_item}
\end{figure}

Furthermore, a glossary of key domain-specific terms is provided\footnote{\url{https://pluto.univie.ac.at/glossary}} to ensure the tool's accessibility to a broad audience.
We also included a dedicated contact option for follow-up queries to ensure users can reach out should the terms used, provided explanations, or recommendations require further clarification.

\subsubsection{Weighting of Questions}

Each item in the questionnaire offers multiple answer choices, with specified minimum and maximum selections per question. Each choice contributes to the risk ($x$-axis) or benefit ($y$-axis) associated with the respondent's data usage. Respondents may also provide an unlisted answer as free text. However, such responses do not affect the final score, as the tool cannot automatically evaluate options not previously encountered. Domain experts manually define and input weights for each answer choice through the Survey Content Management System. For transparency regarding the impact of question weighting on final scores, we publish the weights and their detailed rationale alongside the tool (see \autoref{fig:questionnaire_item} for an example). This information for each questionnaire item is available on the PLUTO website.\footnote{\url{https://pluto.univie.ac.at/weighting}} The respondent's final, visualized score is presented as an $(x, y)$ tuple, where the $x$ component indicates the risk associated with their data use, and the $y$ component represents the associated benefits. To facilitate interpretation, we apply min-max normalization to both components, ensuring a uniform range between $-1$ and $1$. This approach makes the results more accessible and digestible to a broader audience as it supports a common frame of reference~\cite{cleveland1984}. Significantly, this method preserves the relative differences between original values, maintaining the original distribution characteristics among data points. Although the absolute values are adjusted, the underlying patterns remain intact.

\section{Evaluation}
\label{sec:evaluation}

PLUTO's design was refined through a multi-stage evaluation process designed to ensure both its conceptual robustness and practical usability. This process comprised three main phases: (1)~iterative rounds of qualitative feedback on prototypes from domain experts, (2)~an expert workshop, and (3)~a formal summative usability study with non-specialist users to validate the final design.

\subsubsection{Formative Expert Evaluation}

The initial development of PLUTO was guided by two rounds of qualitative feedback (March and August 2023) from 16 domain experts in data science, ethics, law, and governance. This formative feedback focused on improving the tool's clarity, ethical nuance, and usability. Early suggestions led us to clarify the tool’s intended audience (both internal self-assessment and external audits), simplify terminology (e.g., "data user" instead of "applicant"), and add a public glossary and an appendix detailing the question-weighting rationale to improve transparency. Experts also pushed for greater ethical depth, prompting us to rephrase questions to encourage reflection "beyond legal requirements" when considering benefits and risks for marginalized communities. The second round confirmed these improvements and identified a skew in the scoring logic, which caused a disproportionate number of results to be categorized as "high risk, high benefit." In response, we normalized the weighting system to ensure a more balanced distribution of outcomes. Other key improvements included adding context-specific information boxes for each question and implementing a feature to download results as a PDF. A detailed account of this iterative process is available in our online supplement.\footnote{Available at \faGithub~\href{https://github.com/PLUTO-UniWien/PLUTO/blob/main/docs/supplement.pdf}{PLUTO-UniWien/PLUTO/docs/supplement.pdf}.}

\subsubsection{Expert workshop on PLUTO}

Following the official launch of PLUTO at the World Health Summit in October 2023, we hosted a workshop at the Institute for Advanced Study (\textit{Wissenschaftskolleg}) in Berlin with 24 experts (12 in person, 12 online) from the fields of data ethics, data science, and law. Invitees were selected based on prior feedback and purposive selection to ensure diverse perspectives. Participants reviewed PLUTO question by question with the team, raising areas for further refinement.

One concern was that PLUTO seemed biased towards Global North users. 
In response, we revised the question from asking about the user's "track record" of ensuring benefits reached low- and middle-income countries (LMICs) to whether it is "plausible to assume that the data use will benefit people in [LMICs]." We added further clarity within the information box on the relative position of the user. Another key concern raised was that users could gamify PLUTO, particularly as all the weightings are available on the website. Some recommended that the weighting appendix be moved to the end to prevent this issue, though it was decided to retain it at the start for clarity and openness. Instead, we added clarification to the homepage that users were expected to engage with the tool in good faith, and the reliability of results would depend on the answers given. Additionally, we implemented the suggestion to order the multiple-choice answers alphabetically to avoid any perceived bias. This is in line with the conclusions that emerged from extensive discussions concerning the potential gaming of PLUTO. PLUTO has been designed to function as a self-assessment instrument and is not linked to any actionable consequences. Therefore, we do not consider it necessary to provide mechanisms to mitigate the risk of gaming or to make it "gaming-proof". We recognize that this could pose a barrier to the implementation of PLUTO by interested institutions. If regulators wanted to adopt it for their own purposes, it would be necessary to establish suitable auditing mechanisms to ensure maximum veracity and reliability of results. The optimal way to approach this is subject to future work. Feedback also highlighted that certain questions were still complex for users. For example, instead of using "groups protected by anti-discrimination laws," it was decided to change the phrasing to "marginalized groups" while clarifying the legal context in the information box. The inclusion of "beyond formal legal requirements" about protecting marginalized groups from risk was positively received. It was suggested that as the risk from the use of data can never be brought to zero, especially for populations that are already marginalized and underrepresented, it would be more accurate to ask if the data use plausibly presented an "elevated risk" for marginalized communities (as opposed to any risk at all). The results page received positive feedback, with participants highlighting its clarity, accessibility, and the usefulness of the recommendations for improving one's public value score. After the workshop, a summary document was produced, and the team met to discuss and refine the recommendations. A review of all the questions, information boxes, weightings, and results page was conducted and concluded in early 2024.

\subsection{Formal Usability Study}

Following multiple rounds of expert feedback, we conducted a formal usability study to validate PLUTO's design with its intended non-expert audience. The research examined whether first-time users could successfully navigate the tool, apply its question-based framework to a realistic scenario, and accurately interpret the resulting visualization.

\subsubsection{Participants and Procedure}
We recruited $N=12$ participants through both purposive and snowball sampling using university channels and personal networks. Inclusion criteria required participants to have no prior expertise in data governance or significant familiarity with PLUTO and to be fluent in English. The resulting sample, primarily composed of university students from diverse fields including technology, arts, and political science, allowed us to test the tool with a non-specialist audience (a detailed presentation of participants can be seen in our online supplement\footnote{Available at \faGithub~\href{https://github.com/PLUTO-UniWien/PLUTO/blob/main/docs/supplement.pdf}{PLUTO-UniWien/PLUTO/docs/supplement.pdf}.}~\autoref{fig:demographics}). While this sample limits the generalizability of findings, it complements our earlier evaluations which were conducted with a diverse range of experts by focusing on accessibility and ease of use.

Each session was conducted remotely via video conference, lasted approximately 60 minutes, and was recorded with consent. We employed a think-aloud protocol, asking participants to verbalize their thoughts to capture their reasoning, assumptions, and points of confusion in real time. The study comprised four parts: (1)~a scenario-based assessment using the PLUTO tool, (2)~an interpretation of the results visualization, (3)~the collection of quantitative feedback, and (4)~a concluding semi-structured interview.

\subsubsection{Scenario-Based Assessment}
To ground the evaluation, we developed a detailed scenario informed by dialogues with experts and regulators (see \autoref{sec:diagnosys-ai-data-use} in the online supplement). The scenario described ``Diagnosys AI'', a fictional company whose AI-driven diagnostic service represented a challenging, high-risk, low-benefit data use case (Type D). After familiarizing themselves with the scenario, participants used PLUTO to assess it. The focus was on how they used the tool's interface and help features to reason through the assessment. This approach allowed us to evaluate how the tool's design supports a reflective process, rather than testing simple scenario recall.

\subsubsection{Results Interpretation and Comprehension}
Upon submitting the survey, participants were directed to the results page. This part of the session focused on their ability to interpret the visual outputs, including the result type, risk-benefit matrix, numerical scores, and textual recommendations. Moderators used a structured set of probes (see online supplement~\autoref{tab:probes}) to assess comprehension. Participant responses were categorized into one of three outcomes: \textit{Success}, \textit{Eventual Success}, or \textit{Failure}. As shown in online supplement~\autoref{fig:probe_outcomes}, this method revealed a high degree of comprehension across all probes.

\subsubsection{Quantitative Feedback}
Following the interpretation exercise, participants provided quantitative feedback in two stages. First, they rated their experience on a 5-point Likert scale across six dimensions of usability and comprehension (see online supplement~\autoref{tab:ratings}). Second, they completed the System Usability Scale (SUS), a standardized 10-item questionnaire that provides a validated score of perceived usability \cite{brooke1995sus} (see online supplement~\autoref{tab:sus}). These measures are discussed in \hyperref[sec:key-findings]{Key Findings Section}.

\subsubsection{Semi-Structured Interview}
Lastly, a semi-structured interview was conducted to gather deeper insights into the user's experience, identify specific usability barriers, and understand their perception of the tool's overall purpose and intended audience (see online supplement~\autoref{tab:interview} for a list of probes).

\subsubsection{Key Findings}
\label{sec:key-findings}
Quantitative feedback confirmed the tool's high usability. PLUTO achieved a mean System Usability Scale (SUS) score of $91.67$, signifying \textit{best imaginable} perceived usability. The detailed SUS results are shown in online supplement~\autoref{fig:sus_results} and \ref{fig:sus_per_question}. This was corroborated by high ratings across all six subjective measures (see online supplement~\autoref{fig:quant_results}). A thematic analysis of the think-aloud protocols and interview transcripts revealed five core themes explaining the tool's reception: i) agency and empowerment, ii) explainability and trust, iii) learning and reflection, iv) usability and interpretation, and v) visualization comprehension. Representative quotes illustrating these themes are provided in online supplement~\autoref{tab:thematic_quotes}.
The study task prompted learning and reflection, with participants considering nuances they had previously overlooked. As one user stated: 
\begin{quote}
    It makes you think about certain aspects of data use you may haven't considered before... including the more societal impact of data use.
\end{quote}
This educational aspect fostered a sense of agency and empowerment. Users felt the tool equipped them to make informed judgments by structuring a complex ethical task. One participant noted how the tool democratized the assessment:
\begin{quote}
    The one using the tool does not have to make any ethics judgments. They just log in the facts, and the tool basically does the ethical judgment.
\end{quote}
This sense of empowerment was closely tied to the tool’s explainability and was reflected in users’ trust. The textual recommendations were seen as particularly valuable for translating the visual result into a concrete, actionable decision. One user articulated this connection clearly:
\begin{quote}
    If someone didn't use the tool and get the recommendations... they might have given the company their data without knowing the risks. So it's helpful to know the recommendation... and evaluate it for myself again.
\end{quote}

While PLUTO's conceptual framework and final visualization were effective, the study also identified several challenges in usability and interpretation. The most common issues stemmed from terminology. Several users initially assumed ``benefit'' referred to commercial gain for the company, not public value. Similarly, the term ``data user'' caused initial confusion, as some participants (themselves users of the tool) mistook it as referring to them rather than the entity being assessed. Furthermore, we observed a discoverability issue related to UI interaction: some users failed to recognize that underlined terms and info icons throughout the survey provided embedded explanations on hover, perceiving these interactive elements as static text. These findings suggest that clarifying core terminology and improving the visual affordances of interactive features are critical for maximizing the tool's accessibility.

\vspace{-0.75em}
\subsection{Uptake}
Since its launch, we have refined PLUTO based on user feedback. Since early 2024, a user community for PLUTO has been developing, and the team is collaborating with various stakeholders globally who are interested in implementing the tool to make their digital activities more inclusive. These include a non-profit based in Kenya and European public health bodies.

The tool described here is designed to be flexible and adaptable for domain-specific applications. For example, a public health body might need to modify the PLUTO questions, adding, adjusting, or removing specific aspects, to better capture public value in its specific context. Rather than offering a fixed interpretation of the public value in all contexts, we envisage that the tool will act as a useful framework for users to assess and conceptualize the risks and benefits of digital activities in a way that remains adaptable and ultimately grounded in democratic deliberations between data users, policymakers, and the public. As such, the results are not intended for cross-domain comparison. Further, the tool is intended for self-assessment for users who are open and interested in finding out how much public value their intended or current data use (will) provide. Consequently, it will not provide optimal results for users who are not open to self-assessment and reflection. Using PLUTO requires insights on the data user, aspects of benefits and risks of data use, and institutional safeguards--this may not be the case for every person working in a private corporation or public institution interested in the public value of their data use. Optimal results will therefore only be achieved if the user has insight into these various aspects of data use, or can gather them through open communication with others. Organizations fostering secrecy would perhaps not allow access to all the information necessary to optimally use PLUTO. 

\section{Discussion}
The development and evaluation of PLUTO have provided valuable insights into the challenges and opportunities associated with operationalizing a multi-faceted concept like public value. 
A key takeaway was the balance between clarity and inclusivity.  While the tool's purpose and structure were generally well-received, users with different backgrounds would likely require additional guidance to engage effectively.

The inclusion of information boxes and a glossary improved accessibility. They highlighted the necessity of designing tools adaptable to diverse audiences, including those with limited data ethics or analytics expertise. 
While PLUTO may not suit all audiences, this limitation is offset by its design as a prototype that communities or authorities can adapt to their needs and notions of public value.
For example, if a rural community wanted to use PLUTO to assess (and maximise) the public value of a university using data from precision farming for environmental and climate monitoring, the questions and the weightings could be adjusted to better suit this specific setting and context. The iterative process regarding question development that we outline in the \hyperref[sec:background]{Background section} can be used as a basis to adapt PLUTO for different contexts.

A recurring challenge during development was ensuring that PLUTO produced fair and representative results for all users. The initial rounds of feedback revealed concerns about cultural and contextual biases, particularly with regard to the language and framing of questions. These insights led to modifications that have made the tool more inclusive of users from marginalized groups and LMICs. Such adaptations highlight the importance of co-design and collaboration in creating relevant tools across diverse settings.

PLUTO's reception thus far and its growing user community demonstrate the demand for practical tools to assess public value in digital practices. It suggests that PLUTO has the potential to influence data governance at multiple levels, from grassroots initiatives to international policymaking. It will be critical to ensure that PLUTO remains adaptable to different regulatory, cultural, and organizational contexts as its use expands over time. Furthermore, to support collaborative use of PLUTO, future iterations may need to incorporate shared functionalities with differentiated access, enabling multiple stakeholders to use PLUTO to assess collaborative uses of data.



By connecting the concepts of data solidarity and public value and translating quantitative assessments into qualitative insights, PLUTO aims to contribute to a more inclusive digital landscape that addresses the needs of three main audiences: project implementers, policymakers and funders, and the general public.
PLUTO supports individuals and organizations engaged in implementing bottom-up data projects (i.e., any projects that are engaged in the collection, storage, transfer, transformation, and/or analysis of data) by providing them with the means to assess and enhance the public value of their initiatives. 
The tool gives project implementers greater agency to evaluate their data practices without requiring extensive expertise, organize and improve their projects based on public value metrics, and make more informed data usage and sharing decisions. By democratizing evaluative resources, PLUTO enables a broader range of actors to participate meaningfully in the digital ecosystem. 
Policymakers and funding bodies benefit from PLUTO as an instrument for promoting inclusive practices within influential institutions. The tool allows them to assess the public value of data projects, encourage data practices that align with high public value standards, and make informed decisions regarding resource allocation. By providing a framework for evaluating projects, PLUTO helps ensure that funding and policy decisions foster a fairer digital landscape, balancing the benefits and risks of data projects. Finally, PLUTO  aims to empower members of the general public to better understand and assess the public value of data projects that affect them, make informed decisions about personal data sharing (e.g., health data), and participate more actively in discussions and decisions related to the digital ecosystem. This broader engagement amplifies the public’s voice in data-related decisions and fosters a more inclusive digital environment. 

Achieving this inclusivity through a quantitative framework, however, requires careful navigation of inherent complexities. Quantifying the public value of data use is challenging, as reducing complex social phenomena to metrics can oversimplify or misrepresent their true nature \cite{van2014datafication, iliadis2016critical}. This complexity extends to translating results into actionable insights. Visualizations are interpreted subjectively, and their recommendations may have different levels of actionability for users depending on their resources and agency \cite{hullman2019authors}. Consequently, there is an ethical imperative for designers to create visualizations that support clarity while remaining attentive to these interpretive differences, empowering users without imposing a single, definitive judgment \cite{correll2019ethicalvis}.

\section{Conclusion}
In this paper, we introduced PLUTO, a tool that helps stakeholders—including regulators, companies, NGOs, and individuals—assess the public value of data projects by weighing their societal benefits and risks. PLUTO provides a structured, participatory framework that supports more inclusive and accountable decision-making around data use.

We describe its iterative design process, involving visualization prototypes and evaluations with diverse stakeholders, which highlighted both the potential of translating qualitative assessments into quantitative metrics for transparency and the challenges and trade-offs this entails.

PLUTO contributes to greater inclusivity in data governance by making the assessment process more transparent and accessible, though it requires ongoing refinement to ensure diverse perspectives are represented. Its approach aligns with the concept of data solidarity, which shifts focus from individual rights or market interests (e.g., GDPR) to public value, equity, and collective benefit. By operationalizing these principles, PLUTO offers a practical and adaptable tool that embeds ethical evaluation and accountability into digital governance.

\vspace{-1em}
\section{Acknowledgements}
We thank everyone who provided valuable feedback during the design process.
BP and CH gratefully acknowledge support from the Digital Transformations for Health Lab in Geneva (\url{https://dthlab.org/}).

\bibliographystyle{IEEEtran}
\bibliography{references}

@article{reisman2018algorithmic,
  title={{Algorithmic impact assessments: a practical Framework for Public Agency}},
  author={Reisman, Dillon and Schultz, Jason and Crawford, Kate and Whittaker, Meredith},
  journal={AI Now},
  volume={9},
  year={2018}
}

@inproceedings{metcalf2021algorithmic,
  title={{Algorithmic impact assessments and accountability: The co-construction of impacts}},
  author={Metcalf, Jacob and Moss, Emanuel and Watkins, Elizabeth Anne and Singh, Ranjit and Elish, Madeleine Clare},
  booktitle={Proceedings of the 2021 ACM conference on fairness, accountability, and transparency},
  pages={735--746},
  year={2021}
}

@article{kickbusch2021lancet,
  title={{The Lancet and Financial Times Commission on governing health futures 2030: growing up in a digital world}},
  author={Kickbusch, Ilona and Piselli, Dario and Agrawal, Anurag and Balicer, Ran and Banner, Olivia and Adelhardt, Michael and Capobianco, Emanuele and Fabian, Christopher and Gill, Amandeep Singh and Lupton, Deborah and others},
  journal={The Lancet},
  volume={398},
  number={10312},
  pages={1727--1776},
  year={2021},
  publisher={Elsevier}
}

@misc{bryson2014public,
  title={{Public value governance: Moving beyond traditional public administration and the new public management}},
  author={Bryson, John M and Crosby, Barbara C and Bloomberg, Laura},
  journal={Public administration review},
  volume={74},
  number={4},
  pages={445--456},
  year={2014},
  publisher={Wiley Online Library}
}

@article{fukumoto2019public,
  title={{Public values theory: What is missing?}},
  author={Fukumoto, Eriko and Bozeman, Barry},
  journal={The American Review of Public Administration},
  volume={49},
  number={6},
  pages={635--648},
  year={2019},
  publisher={SAGE Publications Sage CA: Los Angeles, CA}
}

@article{prainsack2016thinking,
  title={{Thinking ethical and regulatory frameworks in medicine from the perspective of solidarity on both sides of the Atlantic}},
  author={Prainsack, Barbara and Buyx, Alena},
  journal={Theoretical Medicine and Bioethics},
  volume={37},
  pages={489--501},
  year={2016},
  publisher={Springer}
}

@article{mcmahon2020big,
  title={{Big data governance needs more collective responsibility: the role of harm mitigation in the governance of data use in medicine and beyond}},
  author={McMahon, Aisling and Buyx, Alena and Prainsack, Barbara},
  journal={Medical law review},
  volume={28},
  number={1},
  pages={155--182},
  year={2020},
  publisher={Oxford University Press}
}

@article{hullman2019authors,
  title={{Why authors don't visualize uncertainty}},
  author={Hullman, Jessica},
  journal={IEEE transactions on visualization and computer graphics},
  volume={26},
  number={1},
  pages={130--139},
  year={2019},
  publisher={IEEE}
}

@article{iliadis2016critical,
  title={{Critical data studies: An introduction}},
  author={Iliadis, Andrew and Russo, Federica},
  journal={Big Data \& Society},
  volume={3},
  number={2},
  pages={2053951716674238},
  year={2016},
  publisher={SAGE Publications Sage UK: London, England}
}

@article{van2014datafication,
  title={{Datafication, dataism and dataveillance: Big Data between scientific paradigm and ideology}},
  author={Van Dijck, Jos{\'e}},
  journal={Surveillance \& society},
  volume={12},
  number={2},
  pages={197--208},
  year={2014}
}

@article{cleveland1984,
    author = { William S.   Cleveland  and  Robert   McGill },
    title = {{Graphical Perception: Theory, Experimentation, and Application to the Development of Graphical Methods}},
    journal = {Journal of the American Statistical Association},
    volume = {79},
    number = {387},
    pages = {531-554},
    year  = {1984},
    publisher = {Taylor \& Francis},
    doi = {10.1080/01621459.1984.10478080}
}

@book{Chambers1983,
series = {Wadsworth \& Brooks Cole statistics probability series},
publisher = {Wadsworth \& Brooks Cole Duxbury Press},
isbn = {0871504138},
year = {1983},
title = {{Graphical methods for data analysis}},
language = {eng},
address = {Pacific Grove, Calif. Boston, Mass.},
}

@incollection{shneiderman2003eyes,
  title={{The eyes have it: A task by data type taxonomy for information visualizations}},
  author={Shneiderman, Ben},
  booktitle={The craft of information visualization},
  pages={364--371},
  year={2003},
  publisher={Elsevier}
}

@article{Hogan2025,
author = {Hogan, Connor and Prainsack, Barbara and El-Sayed, Seliem}, 
year = {2025},
month = {1},
title = {{Data solidarity, public value and the future of health data governance. (forthcoming)}},
journal = {De Gruyter Handbook of Digital Health \& Society.}
}

@article{prainsack2022white,
  title={{Data solidarity: a blueprint for governing health futures}},
  author={Prainsack, Barbara and El-Sayed, Seliem and Forg{\'o}, Nikolaus and Szoszkiewicz, {\L}ukasz and Baumer, Philipp},
  journal={The Lancet Digital Health},
  volume={4},
  number={11},
  pages={e773--e774},
  year={2022},
  publisher={Elsevier}
}

@inproceedings{yen2020decipher,
  title={{Decipher: an interactive visualization tool for interpreting unstructured design feedback from multiple providers}},
  author={Yen, Yu-Chun Grace and Kim, Joy O and Bailey, Brian P},
  booktitle={Proceedings of the 2020 CHI Conference on Human Factors in Computing Systems},
  pages={1--13},
  year={2020}
}

@article{crain2023topicvis,
author = {Crain, Patrick and Lee, Jaewook and Yen, Yu-Chun and Kim, Joy and Aiello, Alyssa and Bailey, Brian},
title = {{Visualizing Topics and Opinions Helps Students Interpret Large Collections of Peer Feedback for Creative Projects}},
year = {2023},
issue_date = {June 2023},
publisher = {Association for Computing Machinery},
address = {New York, NY, USA},
volume = {30},
number = {3},
issn = {1073-0516},
url = {https://doi.org/10.1145/3571817},
doi = {10.1145/3571817},
journal = {ACM Trans. Comput.-Hum. Interact.},
month = jun,
articleno = {49},
numpages = {30}
}

@article{brooke1995sus,
author = {Brooke, John},
year = {1995},
month = {11},
pages = {},
title = {{SUS: A quick and dirty usability scale}},
volume = {189},
journal = {Usability Eval. Ind.}
}

@inproceedings{correll2019ethicalvis,
author = {Correll, Michael},
title = {{Ethical Dimensions of Visualization Research}},
year = {2019},
isbn = {9781450359702},
publisher = {Association for Computing Machinery},
address = {New York, NY, USA},
url = {https://doi.org/10.1145/3290605.3300418},
doi = {10.1145/3290605.3300418},
booktitle = {Proceedings of the 2019 CHI Conference on Human Factors in Computing Systems},
pages = {1–13},
numpages = {13},
location = {Glasgow, Scotland Uk},
series = {CHI '19}
}

@article{kennedy2016visconventions,
author = {Kennedy, Helen and Hill, Rosemary and Aiello, Giorgia and Allen, William},
year = {2016},
month = {03},
pages = {1-21},
title = {{The work that visualisation conventions do}},
volume = {19},
journal = {Information, Communication \& Society},
doi = {10.1080/1369118X.2016.1153126}
}

@article{El-Sayed31122025,
author = {Seliem El-Sayed and Ilona Kickbusch and Barbara Prainsack and},
title = {{Data solidarity: Operationalising public value through a digital tool}},
journal = {Global Public Health},
volume = {20},
number = {1},
pages = {2450403},
year = {2025},
publisher = {Taylor \& Francis},
doi = {10.1080/17441692.2025.2450403},
    note ={PMID: 39789994},
URL = {    
        https://doi.org/10.1080/17441692.2025.2450403
},
eprint = {   
        https://doi.org/10.1080/17441692.2025.2450403
}
}

@article{lee2015people,
  title={{How do people make sense of unfamiliar visualizations?: A grounded model of novice's information visualization sensemaking}},
  author={Lee, Sukwon and Kim, Sung-Hee and Hung, Ya-Hsin and Lam, Heidi and Kang, Youn-ah and Yi, Ji Soo},
  journal={IEEE transactions on visualization and computer graphics},
  volume={22},
  number={1},
  pages={499--508},
  year={2015},
  publisher={IEEE}
}

\vspace{-1em}
\section{Author Biographies}

\textbf{Laura Koesten} is an Assistant Professor at MBZUAI, Abu Dhabi, Masdar City, UAE and affiliated with the University of Vienna, 1010, Austria. Her research interests include Human-Data Interaction, Data reuse, and Sensemaking. Koesten received her PhD in Computer Science from the University of Southampton, in collaboration with the Open Data Institute, UK. She is a member of ACM SIGCHI. Contact her at laura.koesten@univie.ac.at.

\textbf{P{\'e}ter Ferenc Gyarmati} is a Master’s Data Science student at the University of Vienna at Vienna, 1010, Austria. His research interests include agentic visual data analysis, visualization recommendation systems and Human-AI interaction. Péter received his BSc in Computer Science from the University of Vienna. Contact him at peter.ferenc.gyarmati@univie.ac.at.

\textbf{Connor Hogan}  is a political and economic research officer based in Dublin. His research interests include public value, data governance and digital transformations in health. Hogan received his Master’s in Political Science from University College Dublin. Contact him at connor.hogan@univie.ac.at. 

\textbf{Bernhard Jordan} is iOS-Developer at WienIT in Vienna, 1030, Austria. His research interests include human-data interaction and conversational user interfaces. Jordan received his MA in Computer Science from the University of Vienna. Contact him at bernhard.jordan@univie.ac.at.

\textbf{Seliem El-Sayed} is an AI ethics and safety researcher in the tech industry. His research interests include AI safety and governance, and public-value-focused approaches to data governance. El-Sayed received his doctorate in Political Science from the University of Vienna. Contact him at seliem.el-sayed@univie.ac.at.

\textbf{Magdalena Eitenberger} is a postdoctoral researcher at the Department of Political Science, University of Vienna at Vienna, 1010, Austria. Her research interests include governance of data and health technologies. Eitenberger received her Dr phil degree in Political Science from the University of Vienna, Austria, and has a background in science and technology studies and communication sciences. Contact her at https://magdalenaeitenberger.com/ or magdalena.eitenberger@univie.ac.at.

\textbf{Marlene, Auer} is a student project assistant at the Department of Political Science at the University of Vienna at Vienna, 1010, Austria. Her research interests include data governance, practices of solidarity and public value of data use. Auer received her BA in International Development from the University of Vienna, Austria. She is a member of Austrian Polar Research Institute (APRI). Contact her at marlene.auer@univie.ac.at.

\textbf{Barbara Prainsack} is professor at the Department of Political Science at the University of Vienna, 1010 Vienna, Austria. Her research interests include data governance and institutions and practices of solidarity. Prainsack received her Dr phil degree in Political Science from the University of Vienna, Austria. She is an elected member of the Academia Europaea (AE), the German National Academy of Science and Engineering (acatech), a corresponding member of the Austrian Academy of Sciences , and a foreign member of the Danish Royal Academy of Sciences and Letters. Contact her at barbara.prainsack@univie.ac.at.

\textbf{Torsten M{\"o}ller} is the head of the Visualization and Data Analysis Research Group, Faculty of Computer Science, Research Network Data Science, University of Vienna, 1090, Vienna, Austria. His research interests include visualization, computer graphics, and data science. Möller received his Ph.D. degree in computer and information science from the Ohio State University, USA. He is a member of the IEEE Computer Society. Contact him at torsten.moeller@univie.ac.at.



\clearpage

\appendix
\section{Supplementary Material}

\subsection{Prototypes}

\begin{figure}[htbp]
    \centering
    \includegraphics[width=0.5\textwidth]{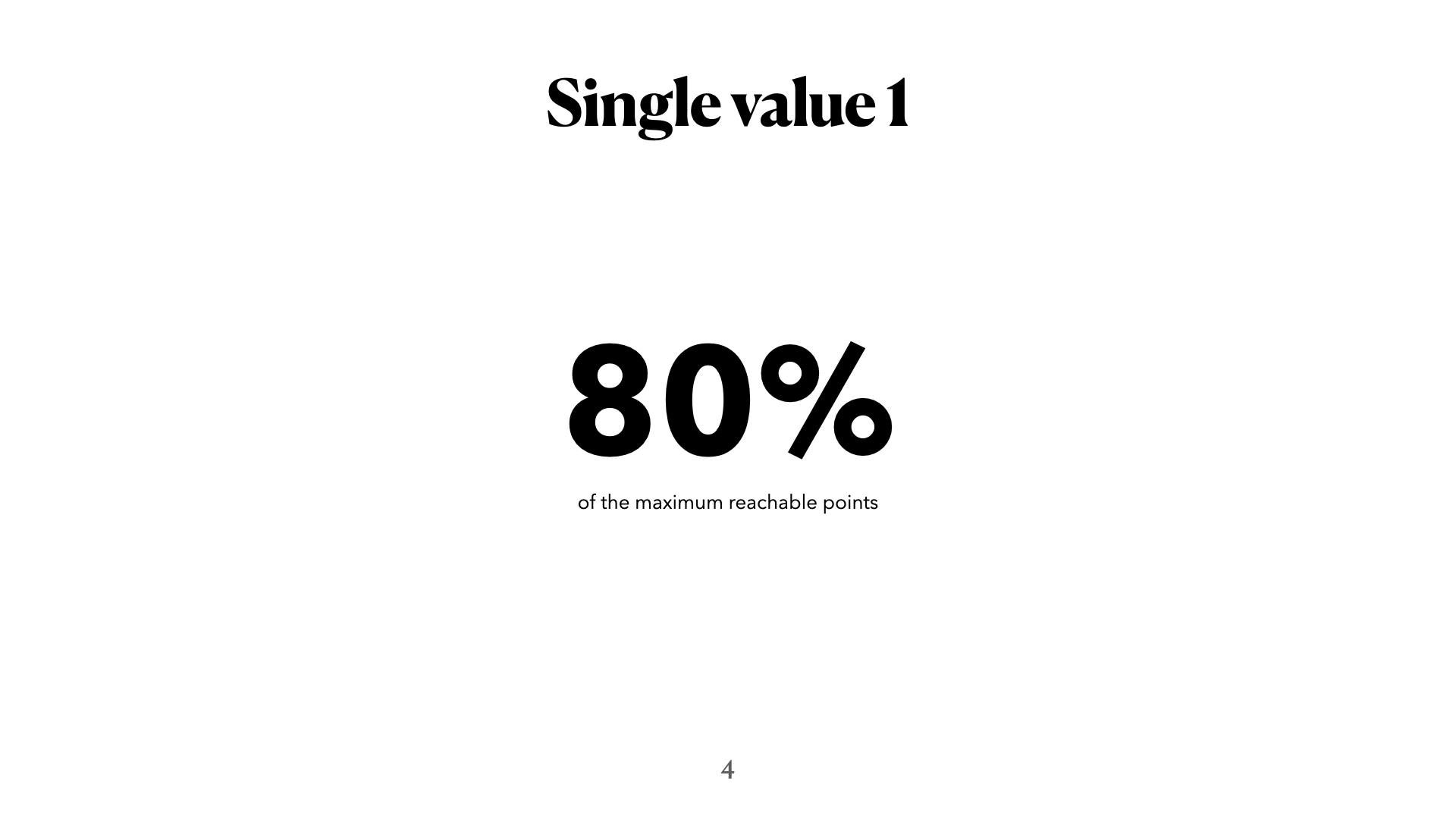}
    \caption{\textbf{Maximum Reachable Points (Single Value 1):} This visualization prominently displays 80\% of the maximum achievable points using a straightforward and bold design. It offers a clear and concise depiction of progress, which is advantageous for quickly assessing performance. However, it does not provide additional context or metrics. By presenting data in an easily understandable format, it aligns with project goals focused on inclusivity and accessibility. While it simplifies numerical complexity and uses an intuitive metric familiar to most audiences, it does not offer the detailed insights required for more in-depth analysis.}
    \label{fig:prototype:singlevalue1}
\end{figure}

\begin{figure}[htbp]
    \centering
    \includegraphics[width=0.5\textwidth]{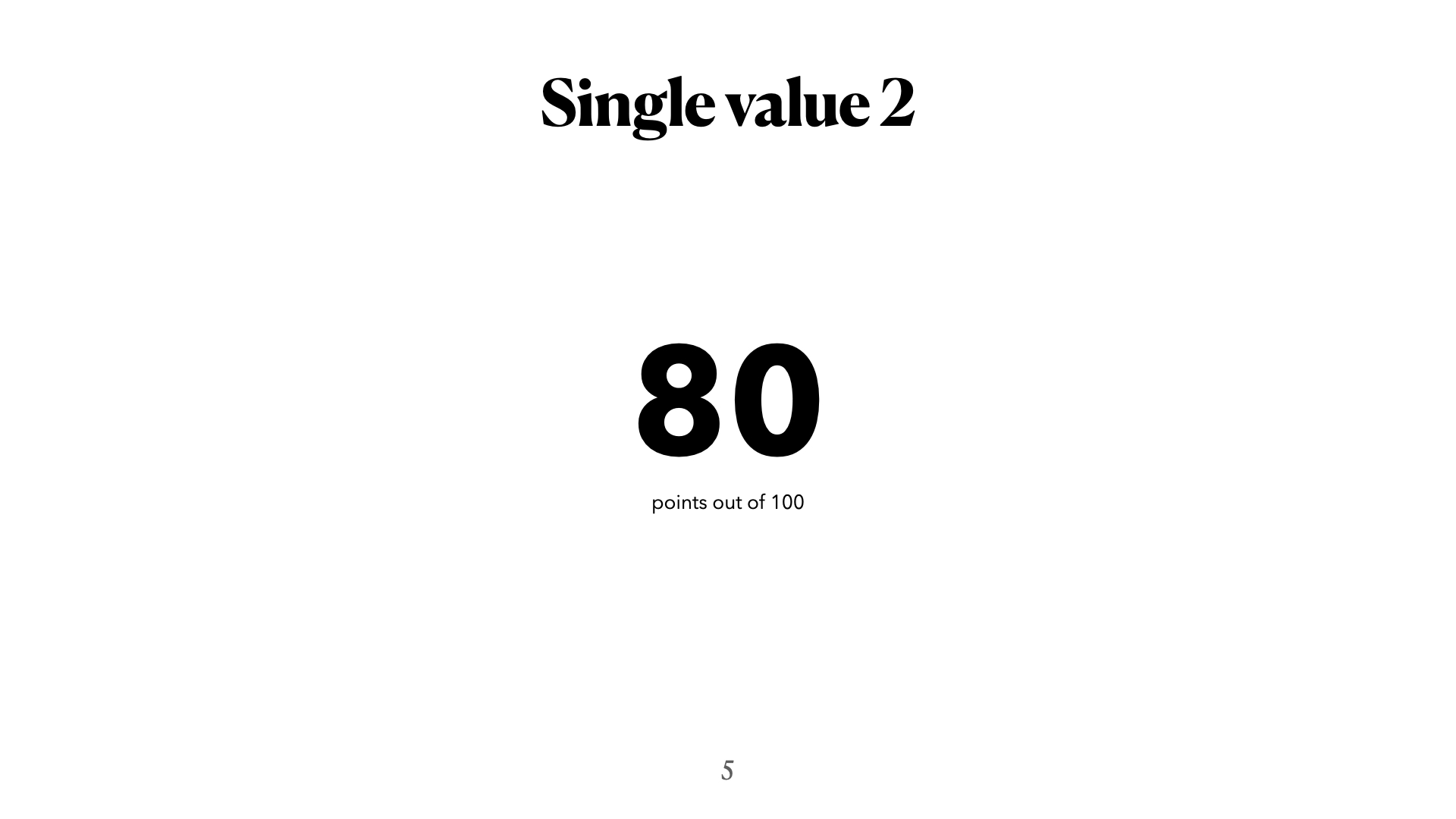}
    \caption{\textbf{Absolute Points (Single Value 2):} This visualization shows a score of 80 points out of a total of 100, offering a straightforward and easy-to-understand representation. It clearly indicates both the achieved score and the total possible points, making it a transparent choice for users. However, it does not facilitate comparison or benchmarking against other scores. By clearly communicating the raw score and total, this design aligns with project goals by providing clarity and simplicity for users who favor direct information. While it reduces ambiguity and builds trust by displaying both the score and total, it lacks context for users less familiar with quantitative assessments, such as whether the score is "good" or "bad."}
    \label{fig:prototype:singlevalue2}
\end{figure}

\begin{figure}[htbp]
    \centering
    \includegraphics[width=0.5\textwidth]{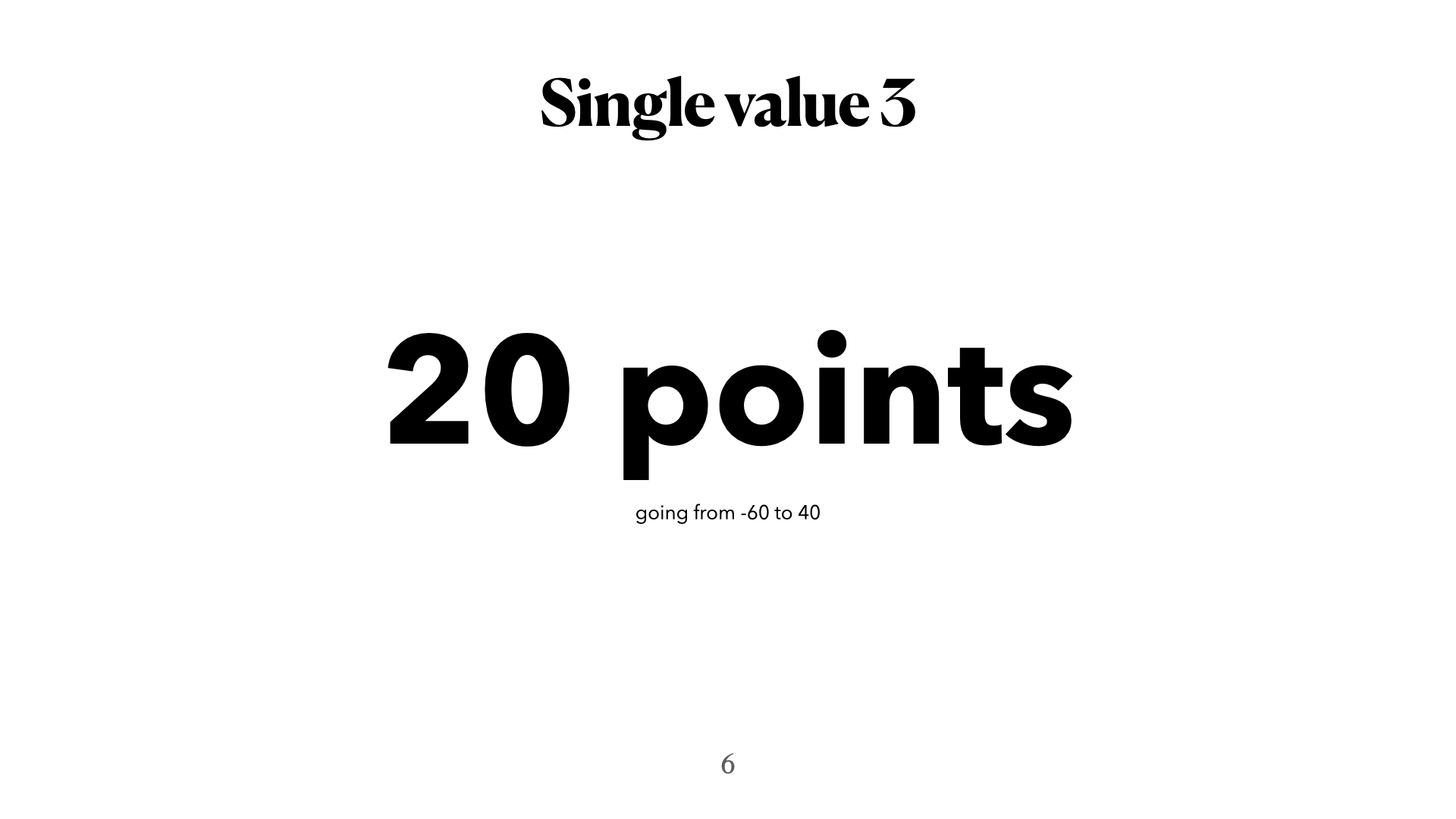}
    \caption{\textbf{Absolute Points on Explicit Scale (Single Value 3):} This visualization highlights a score of 20 points within a specified range from -60 to 40, providing insight into the score's position on a defined scale. While it aids in understanding score relativity, the explicit range might need further explanation for some users, especially due to its skewed nature. By emphasizing scale, it supports users in contextualizing their performance relative to the possible range, aligning with project goals. However, it may present challenges for those less familiar with such scales, affecting its inclusivity.}
    \label{fig:prototype:singlevalue3}
\end{figure}

\begin{figure}[htbp]
    \centering
    \includegraphics[width=0.5\textwidth]{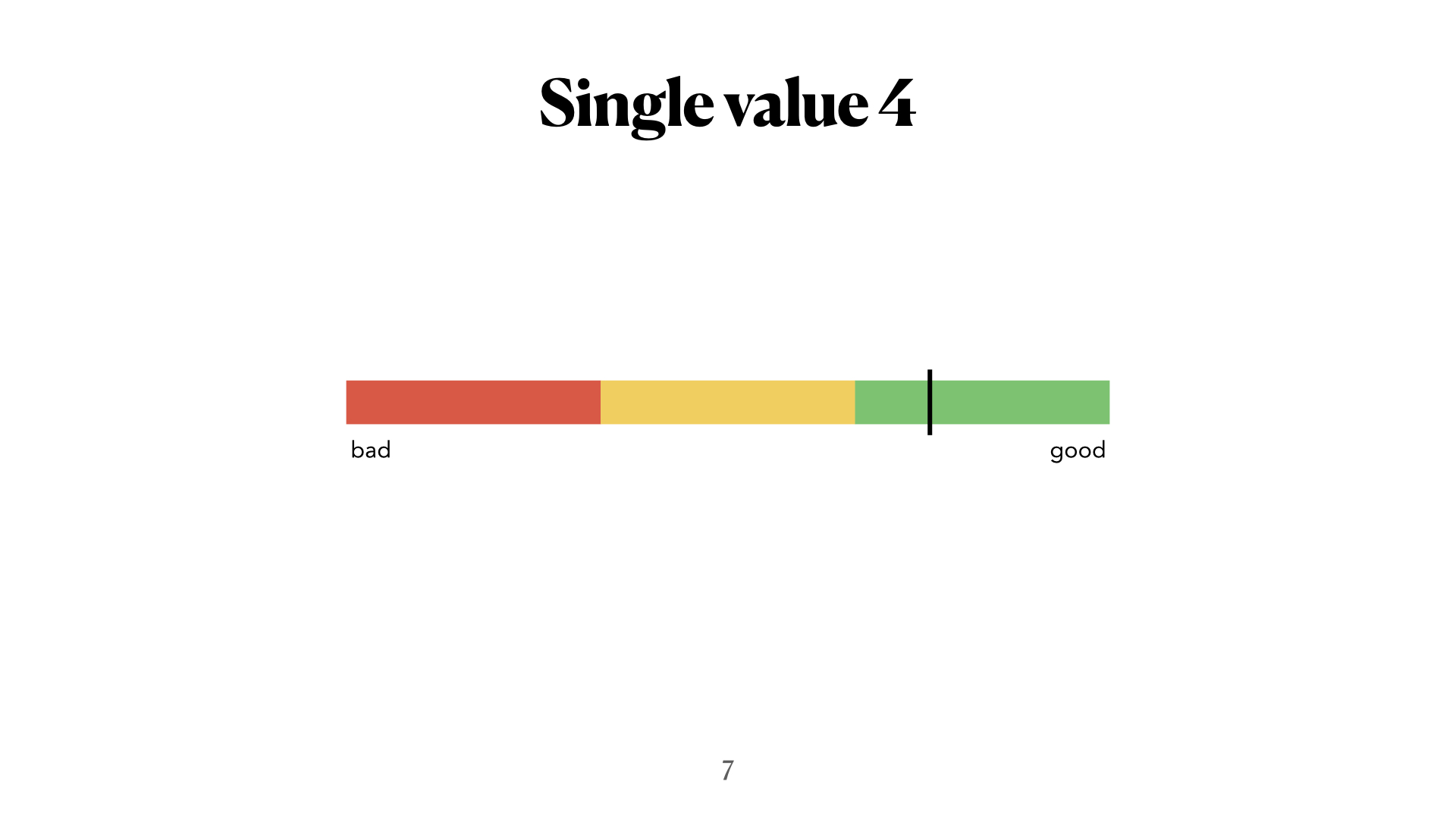}
    \caption{\textbf{Quality Spectrum (Single Value 4):} This figure uses a red-to-green colored bar to depict quality levels, with a marker indicating a position within the "good" range. It leverages intuitive color-coding to communicate quality, aligning with familiar performance indicators. While this method enhances accessibility for users with minimal data literacy, it oversimplifies more complex evaluations, proving to be unsuitable for stakeholders.}
    \label{fig:prototype:singlevalue4}
\end{figure}

\begin{figure}[htbp]
    \centering
    \includegraphics[width=0.5\textwidth]{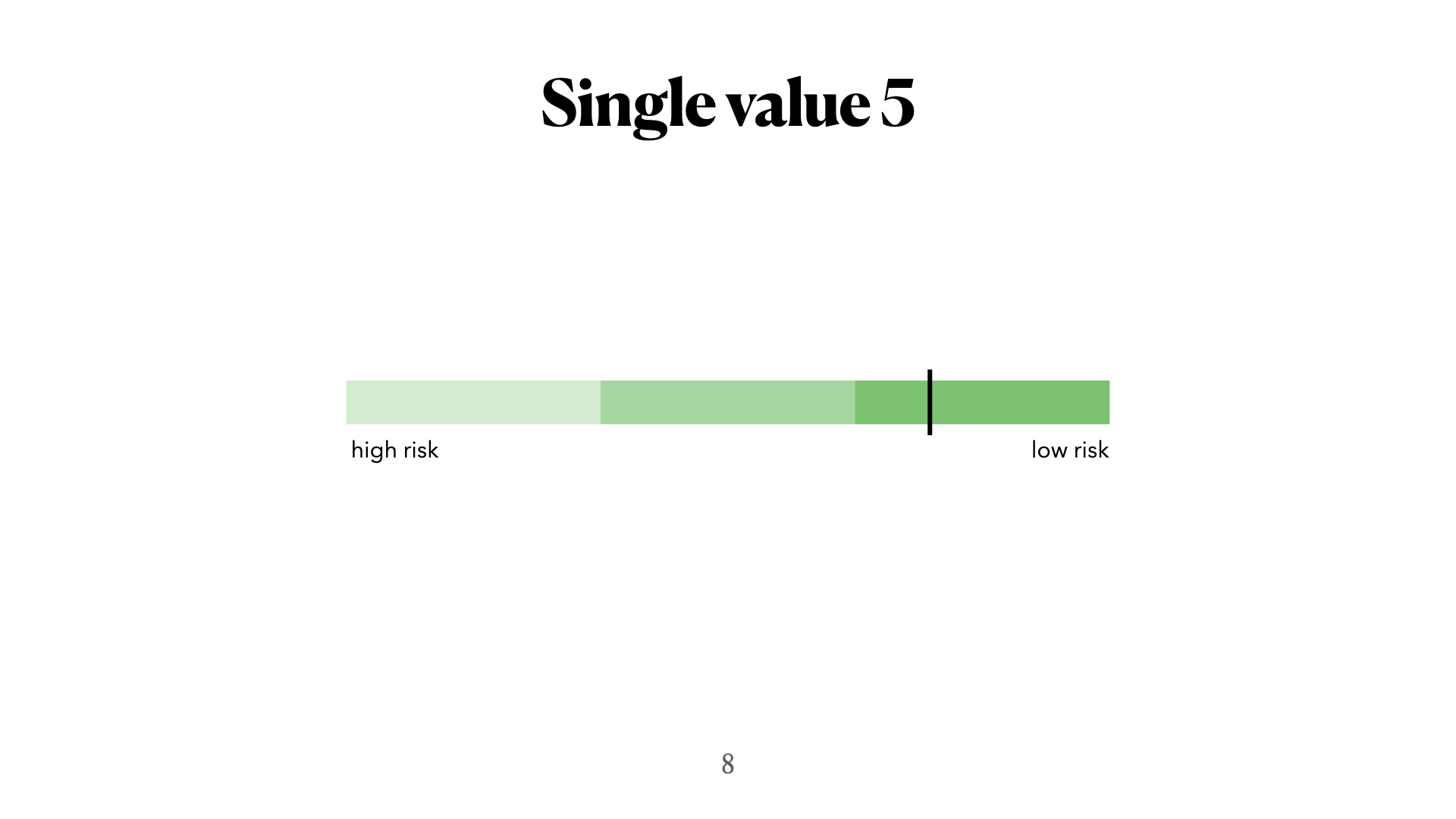}
    \caption{\textbf{Risk Spectrum (Single Value 5):} Depicting risk levels on a gradient from "high risk" to "low risk," this visualization provides a visually intuitive assessment, aided by a color gradient for rapid interpretation. While it aligns with the goal of evaluating data use impact, it lacks precise quantification needed for detailed risk management. The simplicity of the gradient enhances accessibility but might not suffice for users needing precise, actionable guidance.}
    \label{fig:prototype:singlevalue5}
\end{figure}

\begin{figure}[htbp]
    \centering
    \includegraphics[width=0.5\textwidth]{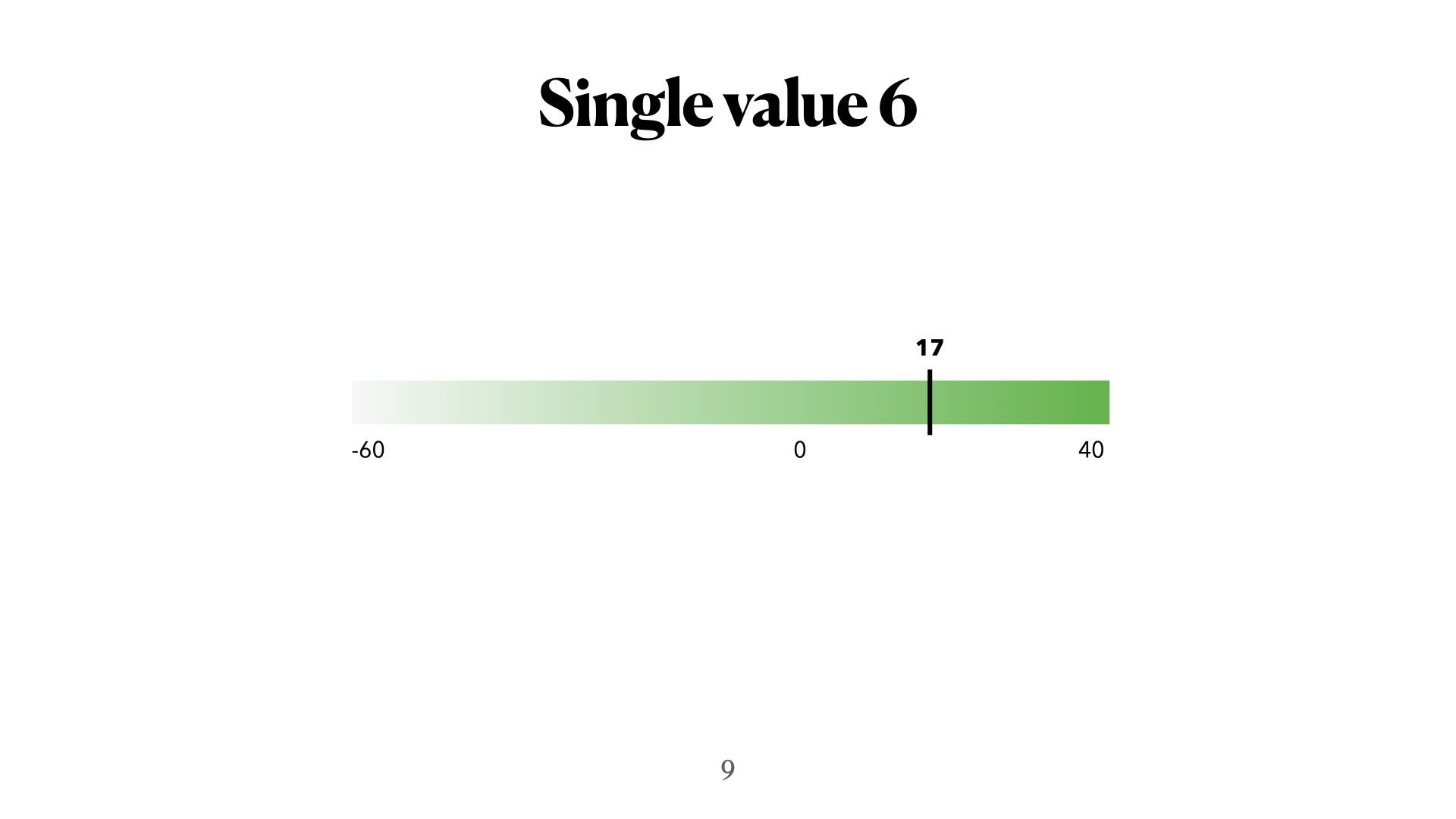}
    \caption{\textbf{Quantified Range (Single Value 6):} Combining a gradient bar with a numerical range from -60 to 40 and a marker at 17, this design merges qualitative and quantitative insights, offering more precise interpretation within a range. While it effectively caters to diverse user preferences by providing both visual and numerical data, the complexity might overwhelm those unfamiliar with such presentations, potentially necessitating additional guidance to enhance inclusivity.}
    \label{fig:prototype:singlevalue6}
\end{figure}

\begin{figure}[htbp]
    \centering
    \includegraphics[width=0.5\textwidth]{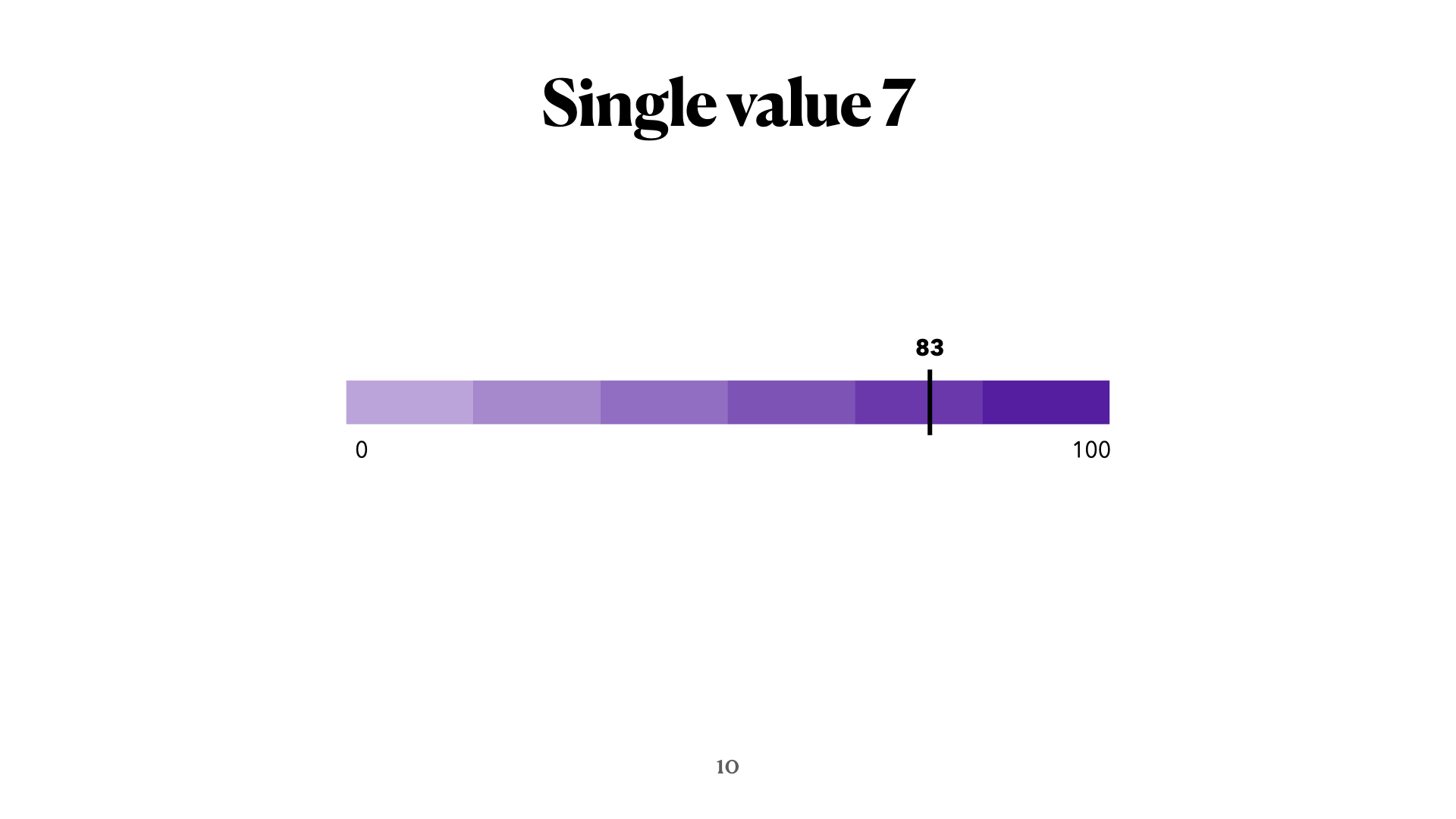}
    \caption{\textbf{Step-wise Gradient (Single Value 7):} This visualization features a gradient bar from light to dark purple, marking a score of 83 out of 100 with discrete steps. It is visually appealing and straightforward, clearly indicating the proportion of the total score achieved. While it excels in clarity with its structured approach, it might lack additional context or benchmarks. This design supports project goals by providing segmented clarity, appealing to users who prefer distinct steps. However, its simplicity may not satisfy advanced users seeking detailed insights.}
    \label{fig:prototype:singlevalue7}
\end{figure}

\begin{figure}[htbp]
    \centering
    \includegraphics[width=0.5\textwidth]{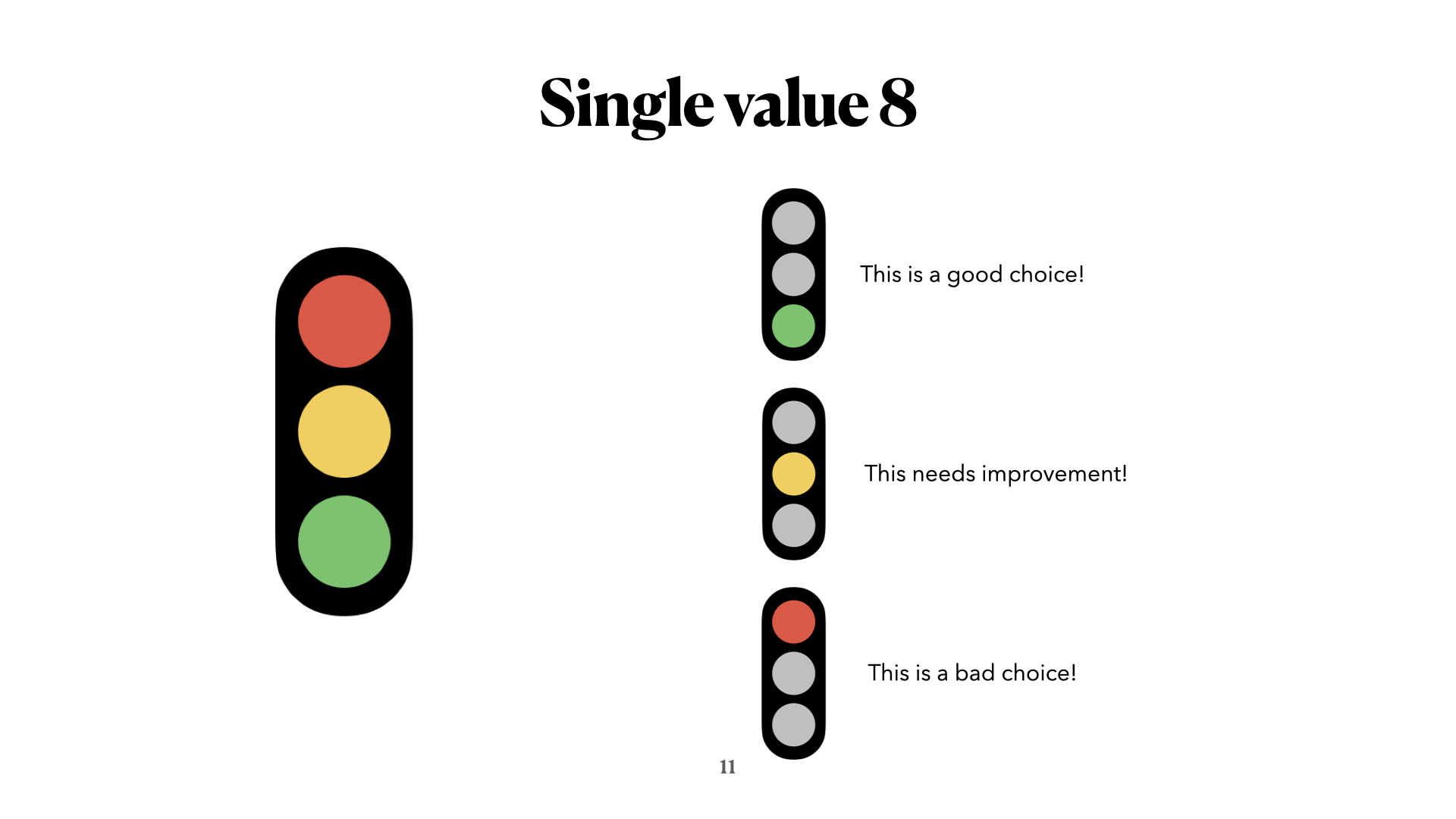}
    \caption{\textbf{Traffic Light Indicator (Single Value 8):} Employing a traffic light metaphor with green, yellow, and red states, this visualization provides a universally recognizable and intuitive evaluation of success or concern. While it simplifies decision-making through categorical grouping, it may lack nuance. By ensuring instant comprehension and minimal numerical demands, it aligns with inclusivity goals, though it may not satisfy users seeking detailed analysis.}
    \label{fig:prototype:singlevalue8}
\end{figure}

\begin{figure}[htbp]
    \centering
    \includegraphics[width=0.5\textwidth]{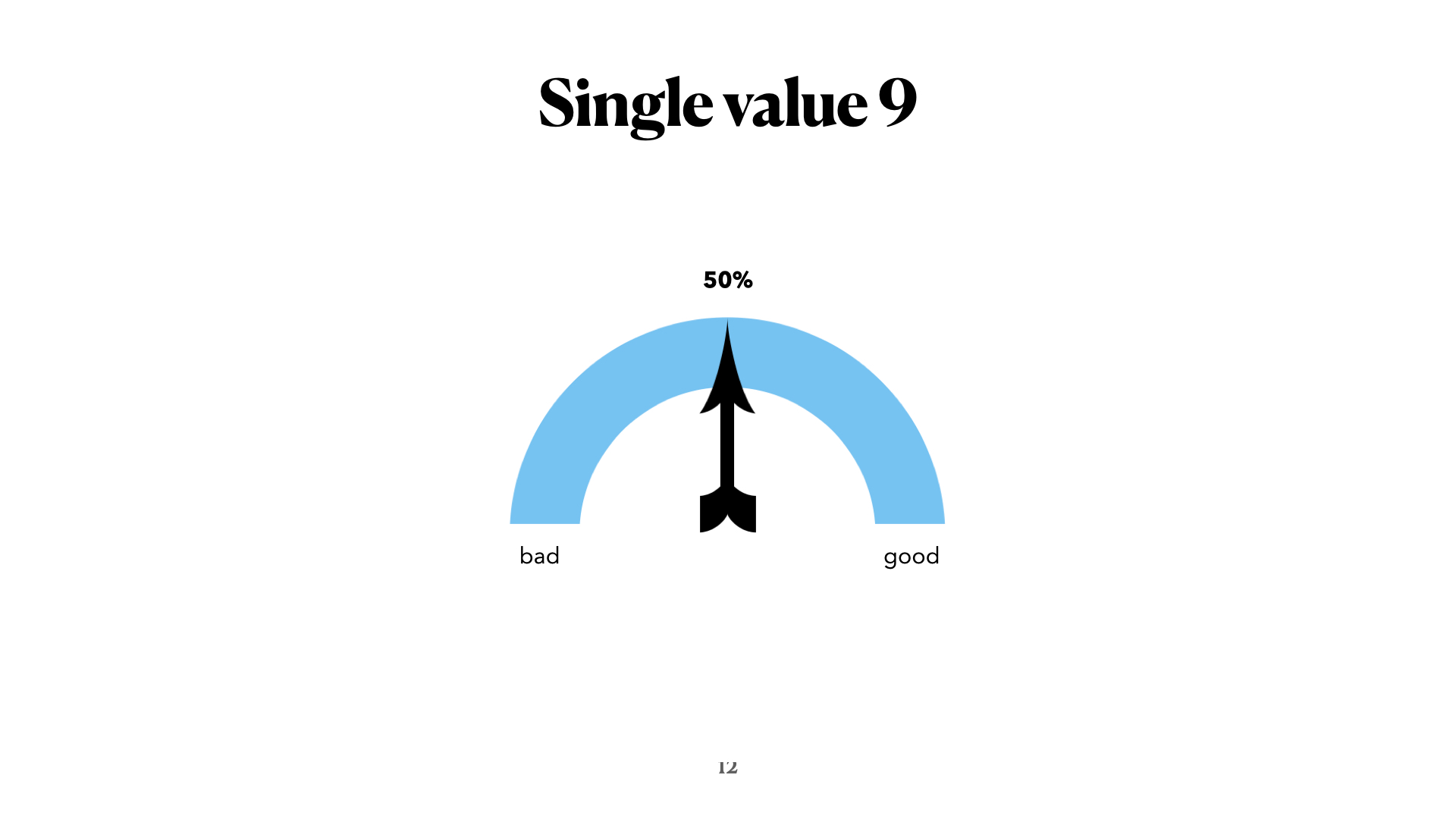}
    \caption{\textbf{Half Arc Indicator (Single Value 9):} Featuring a semi-circular gauge with an arrow at 50\%, labeled from "bad" to "good," this design is visually dynamic and engaging. It effectively represents a balance or midpoint measure, offering an intuitive sense of relative performance. While aesthetically appealing for a wide audience, the exact figures are slightly obscured in this representation.}
    \label{fig:prototype:singlevalue9}
\end{figure}

\begin{figure}[htbp]
    \centering
    \includegraphics[width=0.5\textwidth]{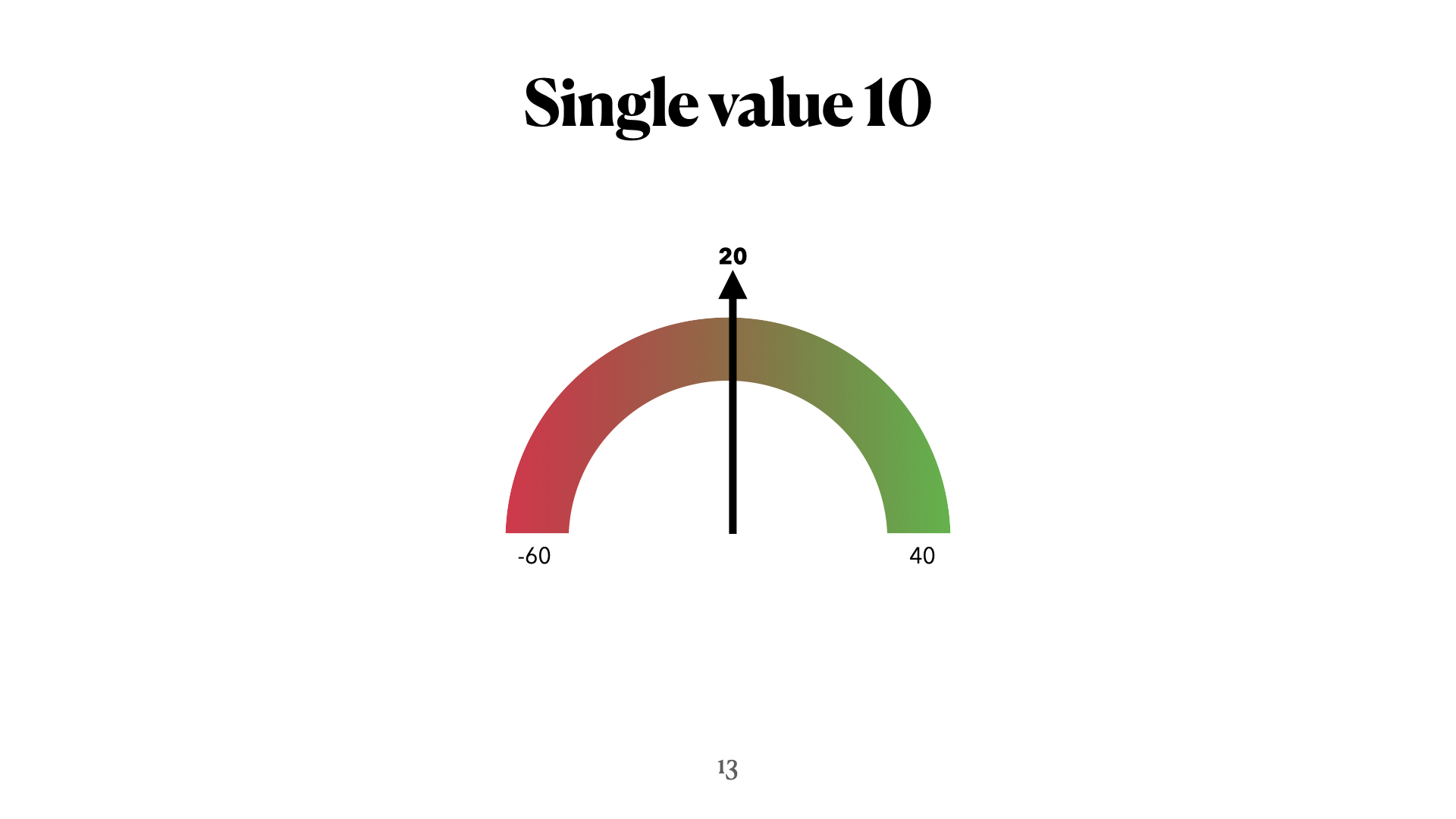}
    \caption{\textbf{Half Arc with Range (Single Value 10):} This visualization combines a semi-circular gauge with a numerical range of -60 to 40, marked at 20. It merges numerical detail with a visual indicator, intuitive for assessing performance within a defined range. While it balances visual appeal with information, interpretation might need further explanation for less data-literate users, especially regarding the range's significance.}
    \label{fig:prototype:singlevalue10}
\end{figure}

\begin{figure}[htbp]
    \centering
    \includegraphics[width=0.5\textwidth]{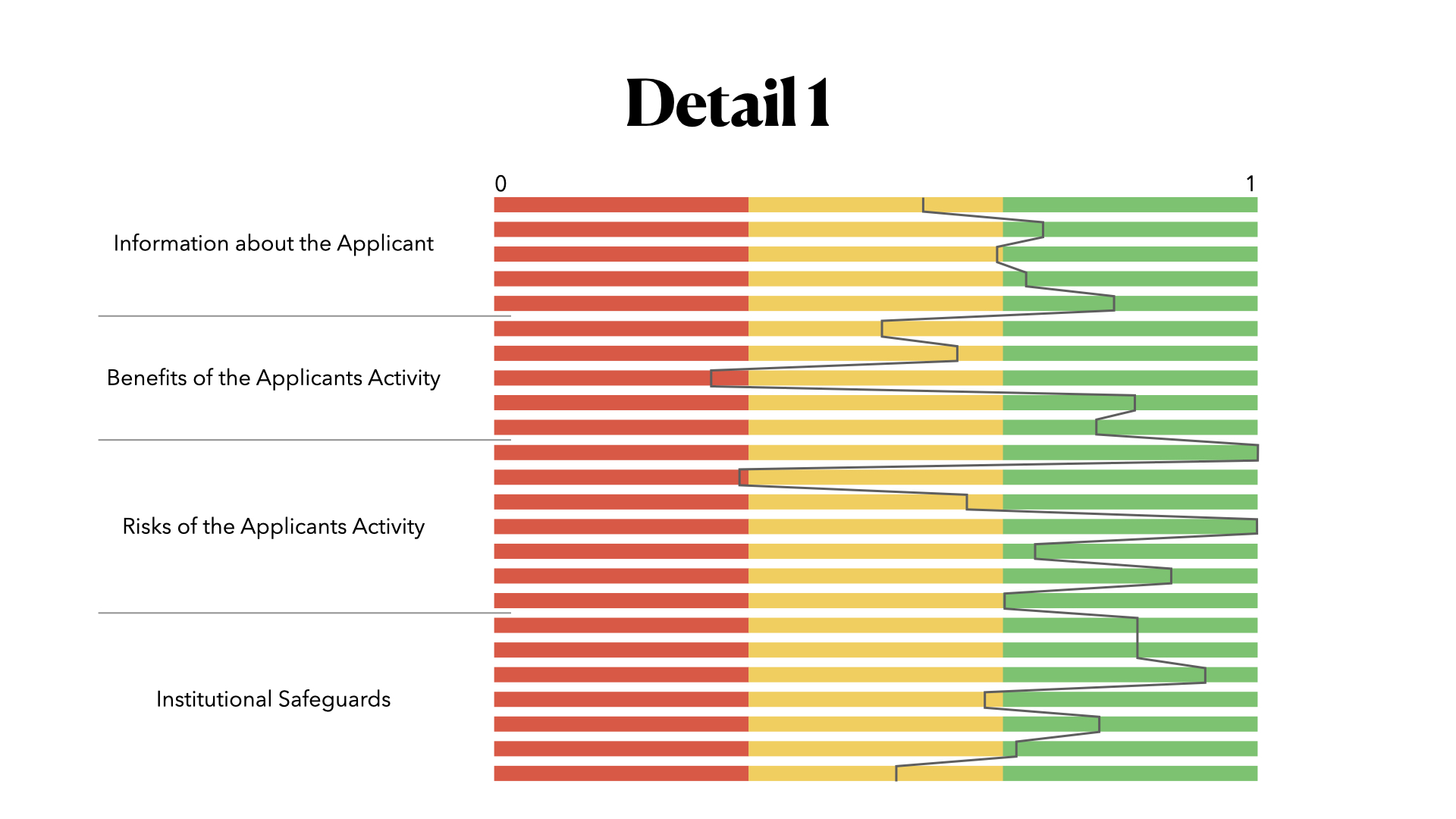}
    \caption{\textbf{Gradient Horizontal Bars with Line Overlay (Detail 1):} This visualization features horizontal bars segmented into red, yellow, and green to represent negative, neutral, and positive results, with a line graph overlay illustrating trends. It highlights variations and categories, offering actionable insights. However, the line overlay may complicate interpretation for some users. The design aligns with project goals by exploring relationships among dimensions and requires careful attention to ensure clarity for all users.}
    \label{fig:prototype:detail1}
\end{figure}

\begin{figure}[htbp]
    \centering
    \includegraphics[width=0.5\textwidth]{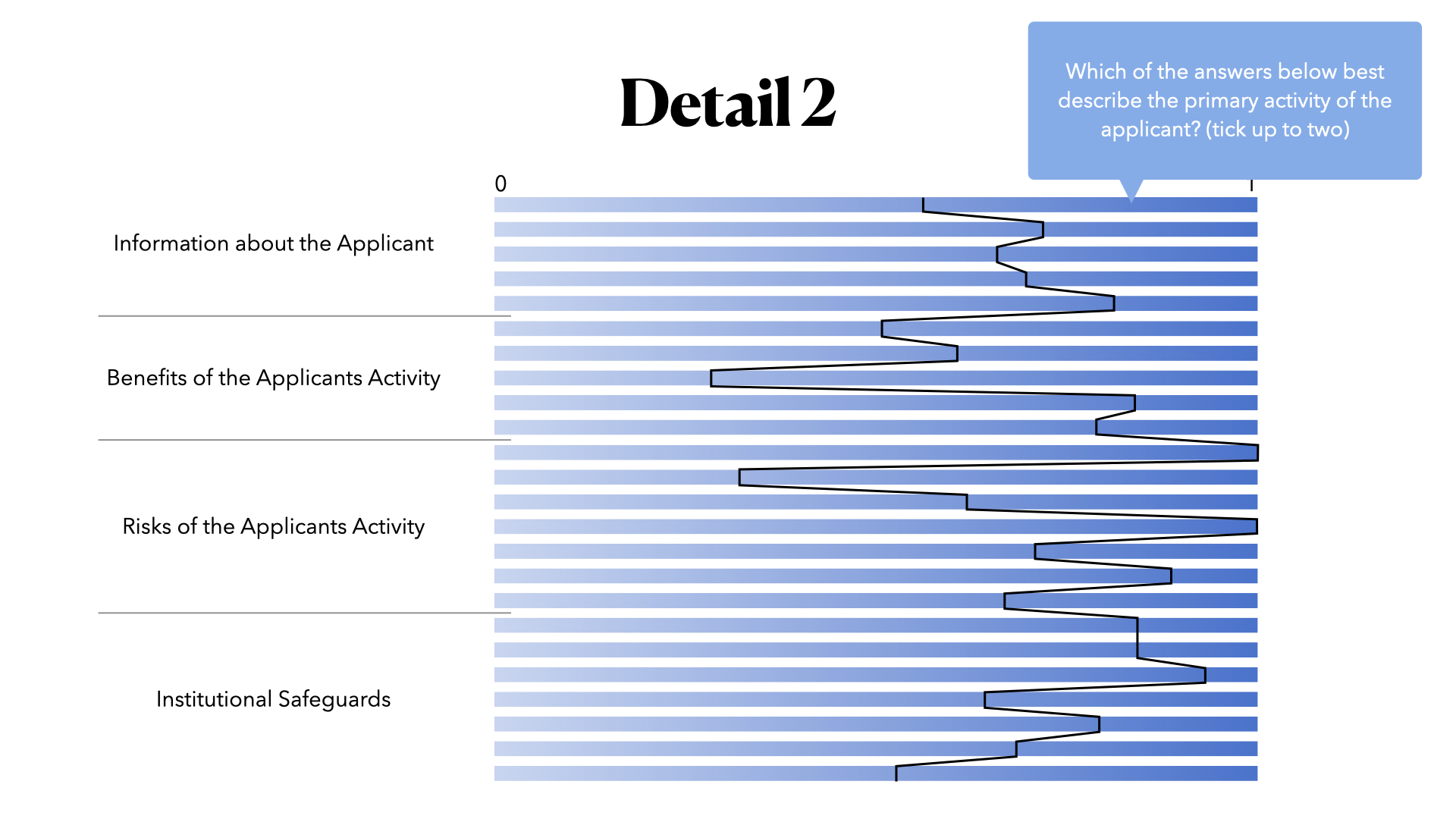}
    \caption{\textbf{Gradient Bars with Prompt Overlay (Detail 2):} Similar to \autoref{fig:prototype:detail1}, this chart uses a blue gradient background and includes a textual prompt for guidance. It enhances interpretability with context but may rely too heavily on the gradient, potentially overshadowing detailed insights. By providing explanatory prompts, it bridges understanding gaps and aligns with inclusive goals, though supplementary explanations might be necessary for complete clarity.}
    \label{fig:prototype:detail2}
\end{figure}

\begin{figure}[htbp]
    \centering
    \includegraphics[width=0.5\textwidth]{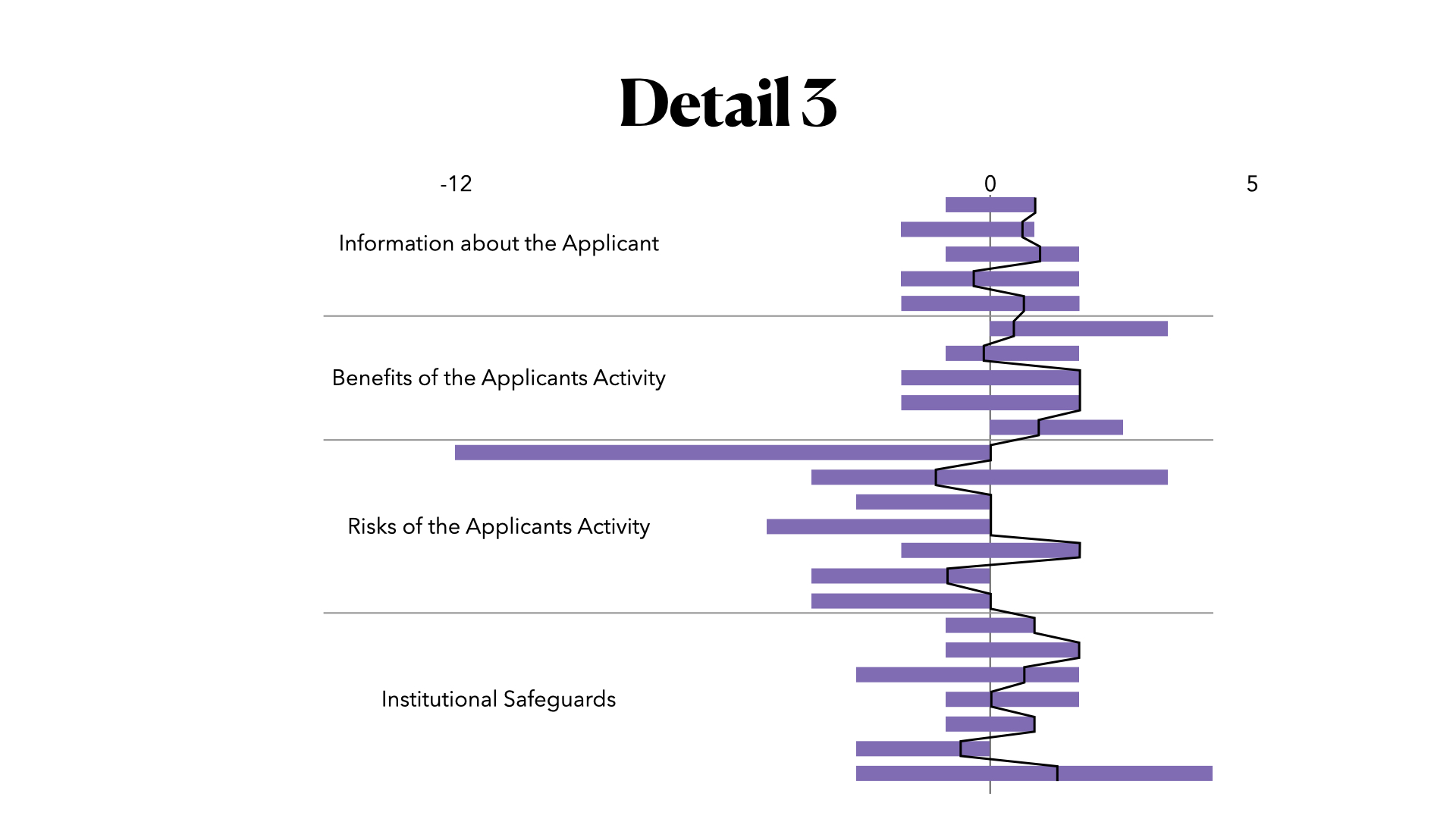}
    \caption{\textbf{Horizontal Bars with Diverging Axes (Detail 3 - Type 1):} This chart presents horizontal bars diverging from a central axis, with an overlaying line graph to show trends. It clearly distinguishes between negative and positive metrics but may increase cognitive load for users unfamiliar with this format. The design supports nuanced exploration of imbalances, aligning with project goals, but additional labeling or legends might be needed to enhance understanding.}
    \label{fig:prototype:detail3}
\end{figure}

\begin{figure}[htbp]
    \centering
    \includegraphics[width=0.5\textwidth]{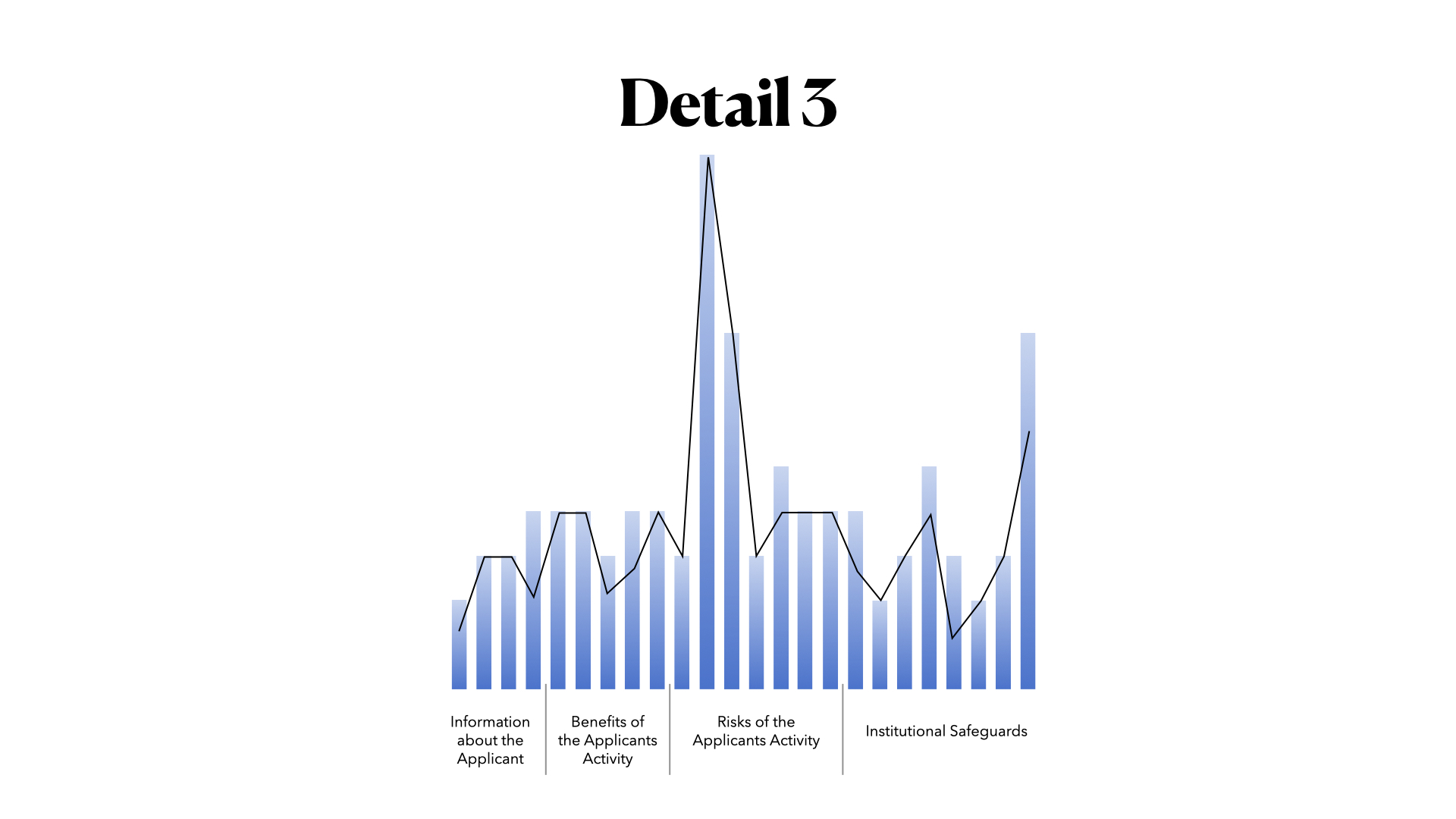}
    \caption{\textbf{Vertical Bar Chart with Line Overlay (Detail 3 - Type 2):} Featuring vertical bars for each category with a line graph overlay, this design simplifies comparison and adds trend-based analysis. While effective for longitudinal assessments, it may become visually dense. The familiar bar structure ensures accessibility, though simplified annotations might help prevent overwhelming users.}
    \label{fig:prototype:detail4}
\end{figure}

\begin{figure}[htbp]
    \centering
    \includegraphics[width=0.5\textwidth]{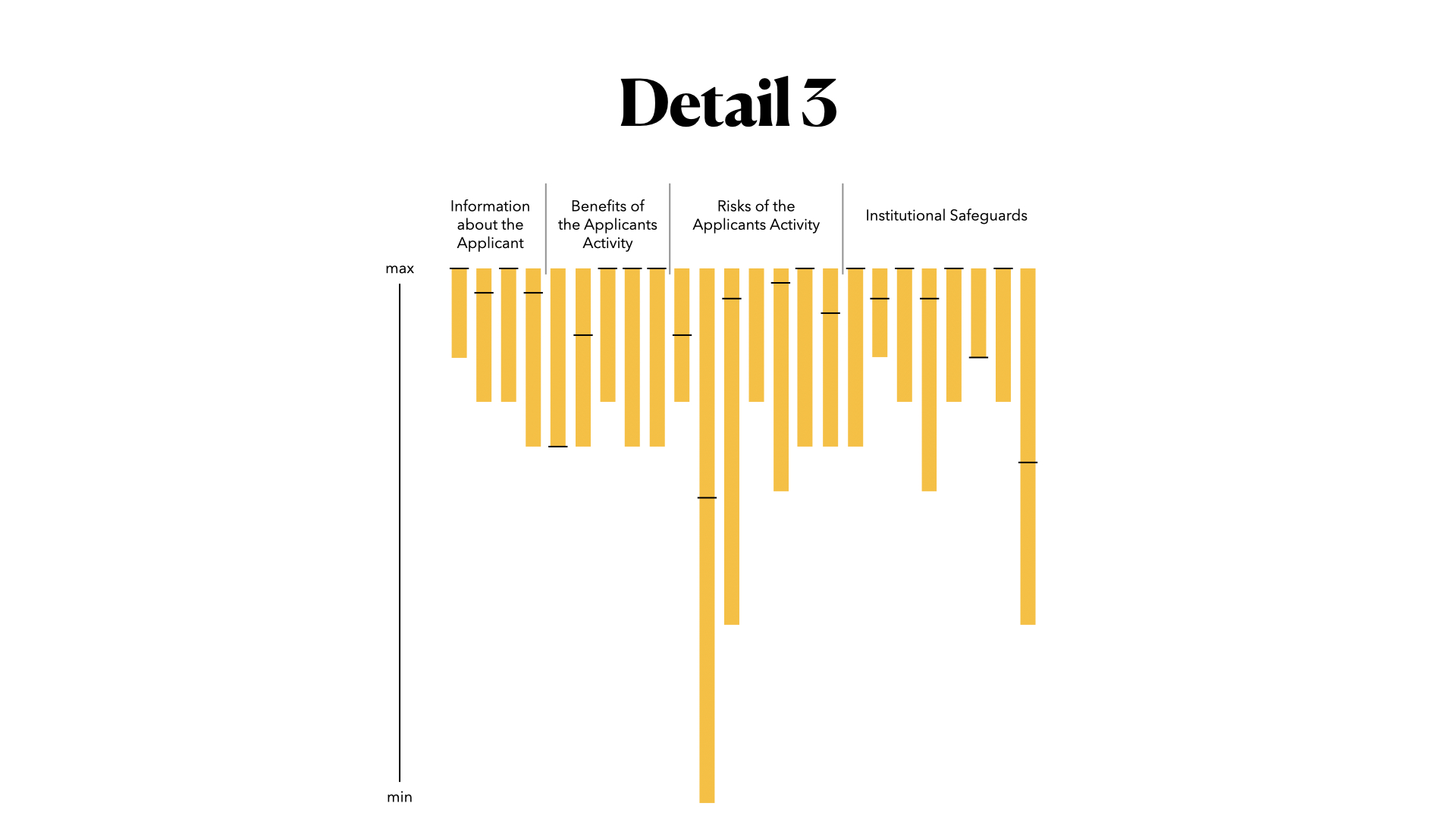}
    \caption{\textbf{Inverted Bars with Max/Min Indicators (Detail 3 - Type 3):} This visualization uses vertical bars extending downward from a central axis, marking maximum and minimum values. It emphasizes the range of values but may confuse users unfamiliar with inverted scales. By highlighting extremes, it supports identifying critical thresholds and risks, aligning with project goals, though annotations may be necessary to guide interpretation.}
    \label{fig:prototype:detail5}
\end{figure}

\begin{figure}[htbp]
    \centering
    \includegraphics[width=0.5\textwidth]{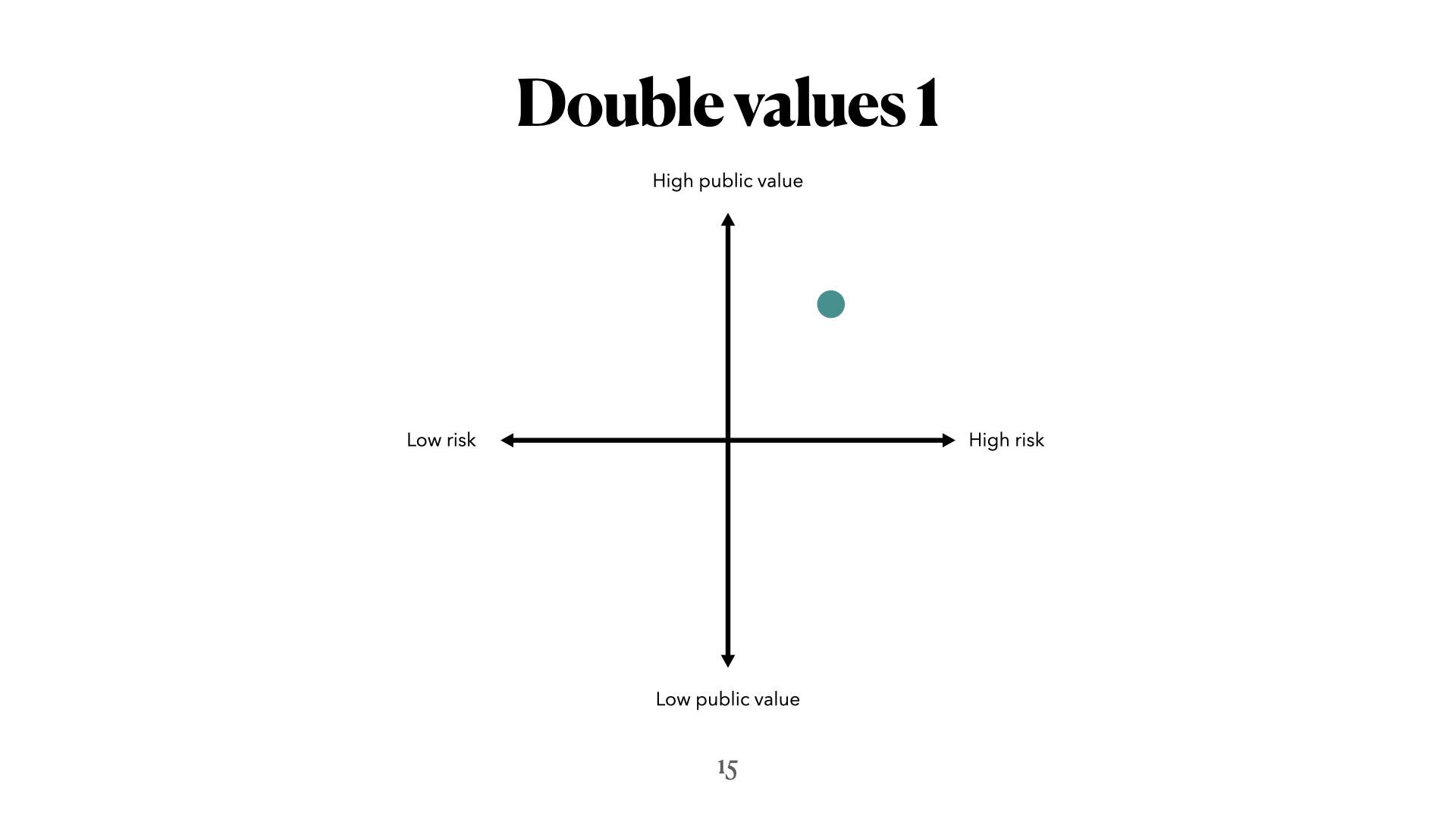}
    \caption{\textbf{Basic 2D Grid (Double Values 1):} This simple Cartesian plane is divided into quadrants with axes representing "Public Value" and "Risk," plotting a single data point to indicate the respondent's result. The straightforward design is easy to interpret, offering clear categorization (e.g., high public value and high risk). While lacking in contextual elements like colors or ranges, its minimalism ensures clarity and accessibility for those with lower data literacy, aligning with inclusivity goals by avoiding complex visuals. However, it may not provide deeper insights into trends or distributions.}
    \label{fig:prototype:doublevalues1}
\end{figure}

\begin{figure}[htbp]
    \centering
    \includegraphics[width=0.5\textwidth]{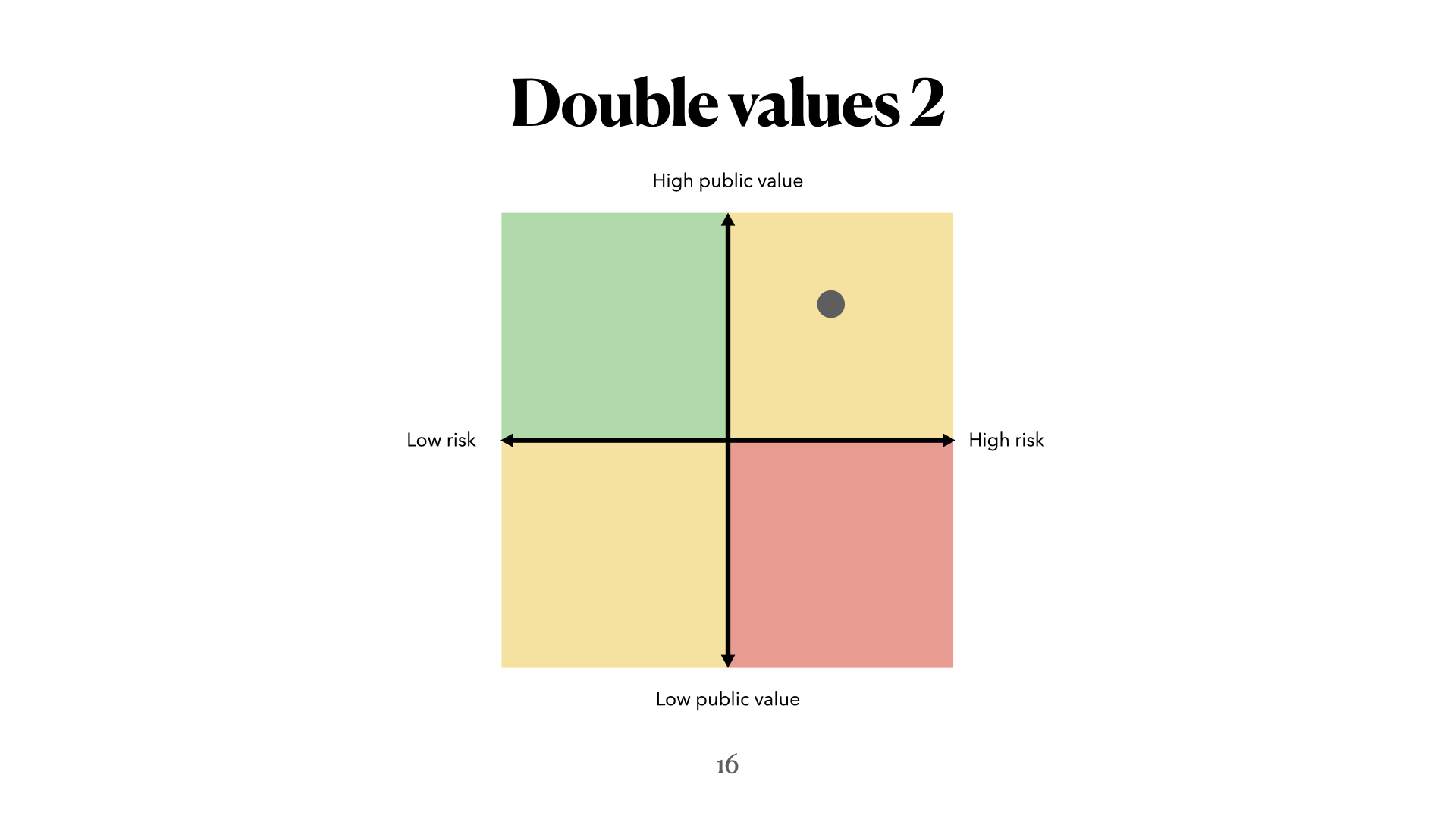}
    \caption{\textbf{Quadrants with Color Coding (Double Values 2):} Building on the basic 2D grid, this design incorporates color coding for each quadrant, such as green for "Low Risk/High Public Value" and red for "High Risk/Low Public Value." This enhances interpretability and offers immediate visual feedback, helping categorize results into favorable or unfavorable zones. While intuitive for most users, the design may not be accessible for colorblind users without alternative indicators, and it might oversimplify nuanced relationships between public value and risk. The design illustrated by this prototype was ultimately chosen for the final version of the tool in combination with complementary visual components to balance out the discussed shortcomings.}
    \label{fig:prototype:doublevalues2}
\end{figure}

\begin{figure}[htbp]
    \centering
    \includegraphics[width=0.5\textwidth]{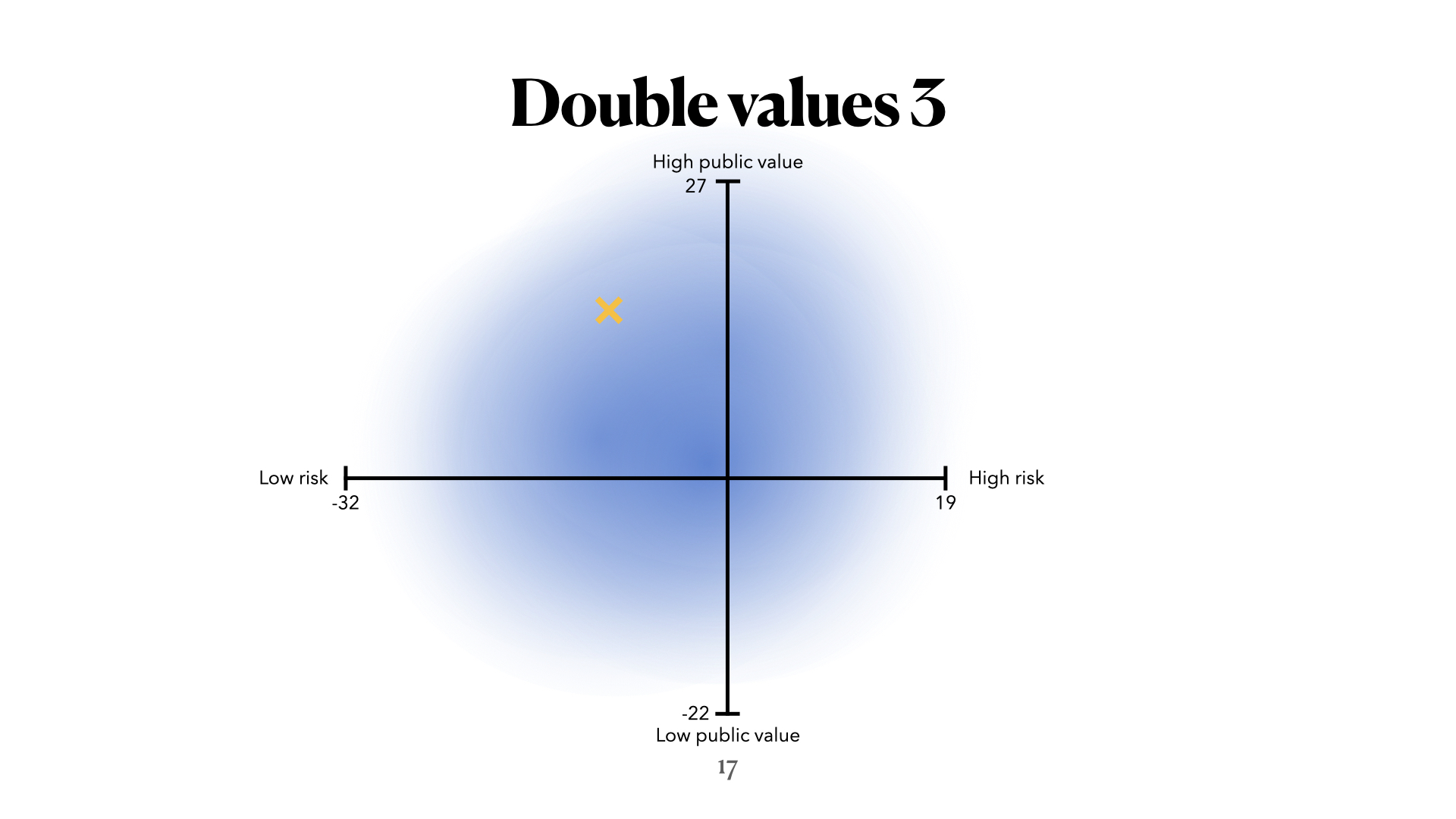}
    \caption{\textbf{Heatmap Overlay (Double Values 3):} This visualization merges a Cartesian plane with a heatmap, using a background color gradient to indicate data density, with a yellow "X" representing the respondent's position. It adds richness by highlighting trends or clusters, providing context on how a result compares to others. While visually engaging, it may overwhelm non-technical audiences, requiring supplementary guidance for full inclusivity. For data-savvy users, however, the additional context enhances interpretability and engagement.}
    \label{fig:prototype:doublevalues3}
\end{figure}

\begin{figure}[htbp]
    \centering
    \includegraphics[width=0.5\textwidth]{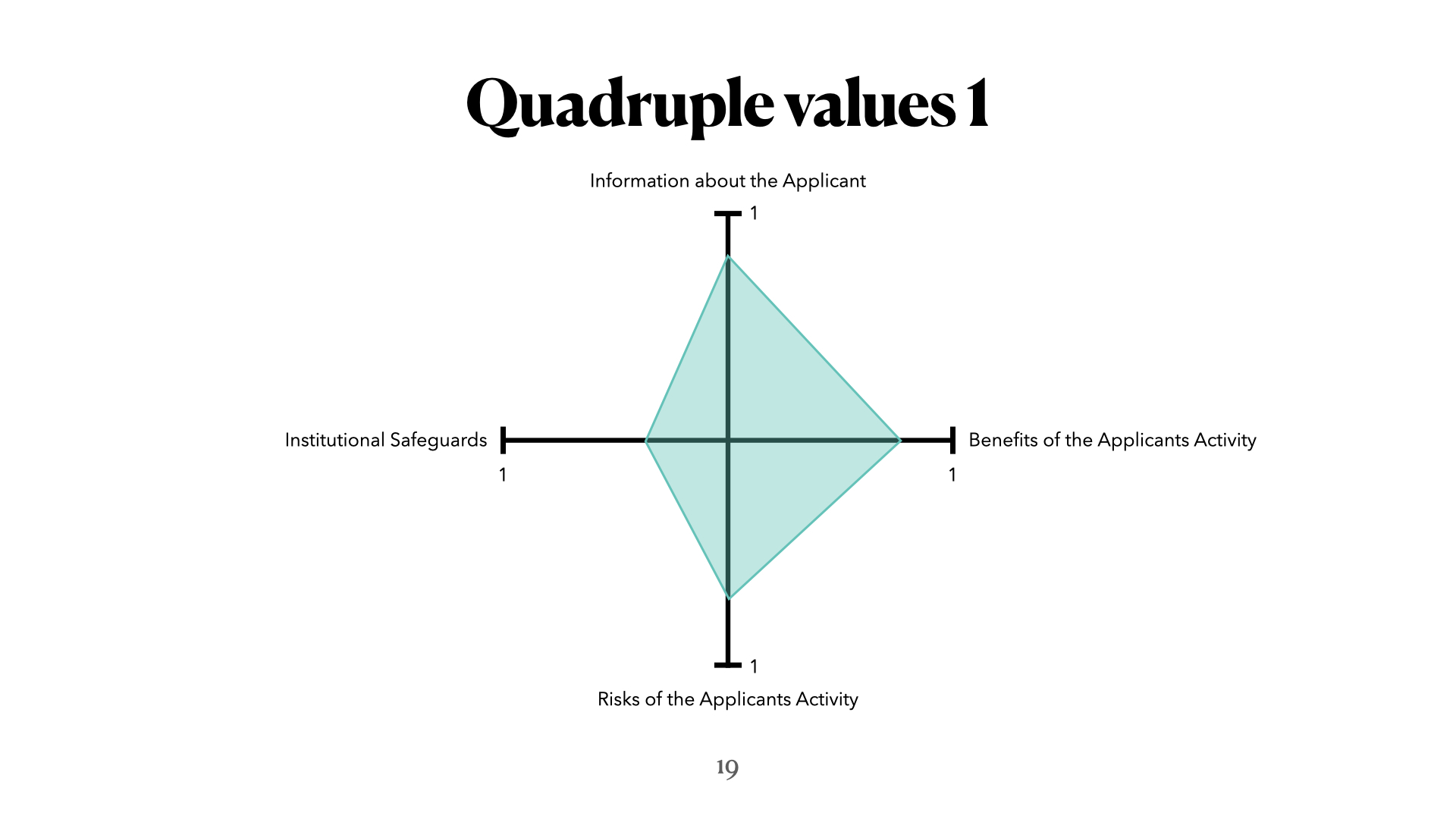}
    \caption{\textbf{Radial Diagram (Quadruple Values 1):} This radial chart connects values across four axes to form a polygon, intuitively visualizing balance or disparities between dimensions. It highlights strengths and weaknesses effectively, although it may be challenging for those unfamiliar with radial charts. By conveying a multidimensional perspective, it aligns with project goals of holistic and inclusive data interpretation. While engaging for varied cognitive preferences, unfamiliar users may benefit from annotations or a legend to aid understanding.}
    \label{fig:prototype:quadruplevalues1}
\end{figure}

\begin{figure}[htbp]
    \centering
    \includegraphics[width=0.5\textwidth]{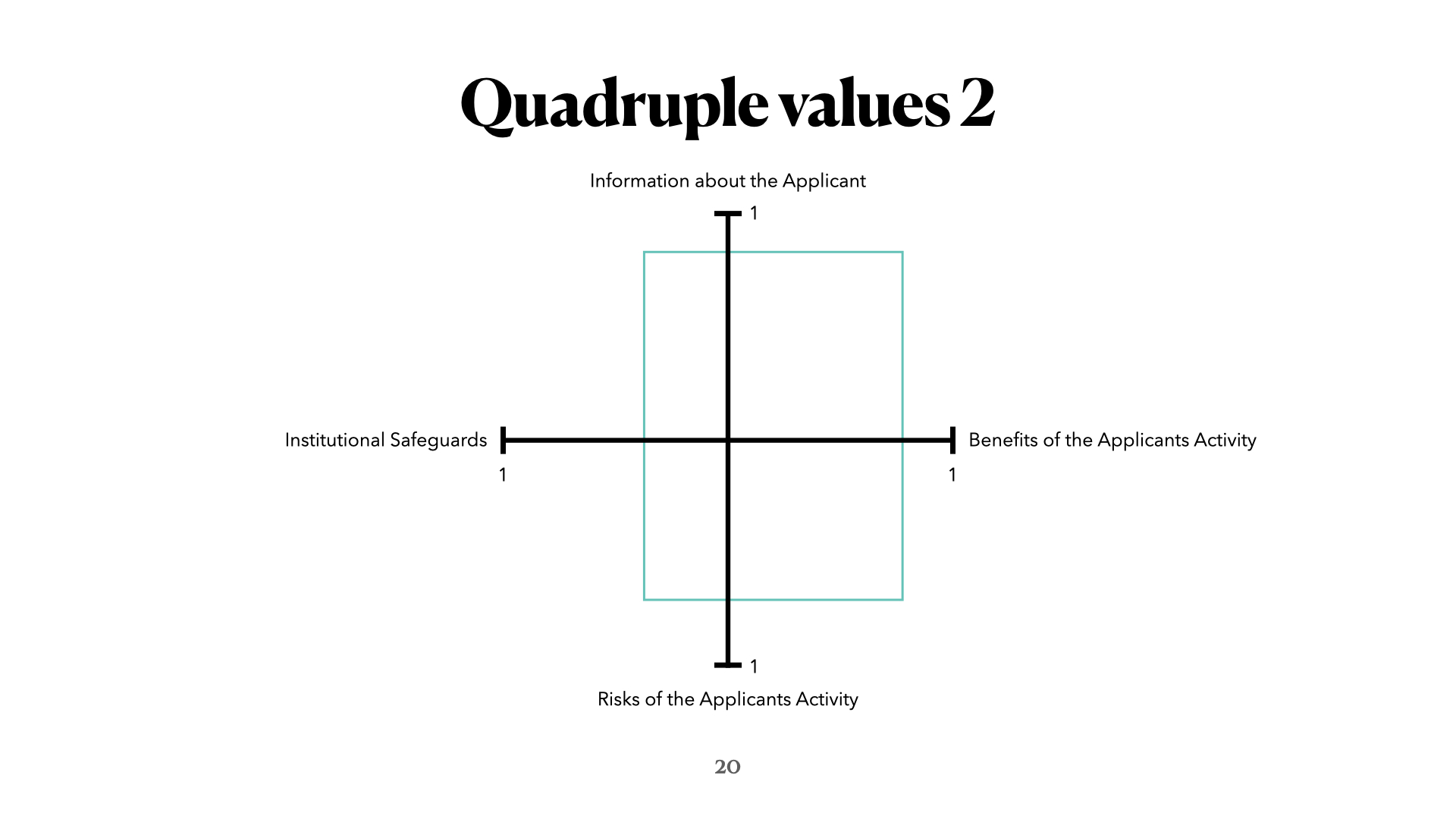}
    \caption{\textbf{Rectangular Overlay (Quadruple Values 2):} This design features four axes radiating from a central point to form a rectangle, indicating clear boundaries for maximum and minimum values. Simpler than radial charts, it reduces cognitive load but may lack proportional representation. It effectively communicates data limits and comparative utility, aligning with simplicity and usability goals. The minimalistic approach enhances accessibility, appealing to diverse visual literacy levels without overwhelming complexity.}
    \label{fig:prototype:quadruplevalues2}
\end{figure}

\begin{figure}[htbp]
    \centering
    \includegraphics[width=0.5\textwidth]{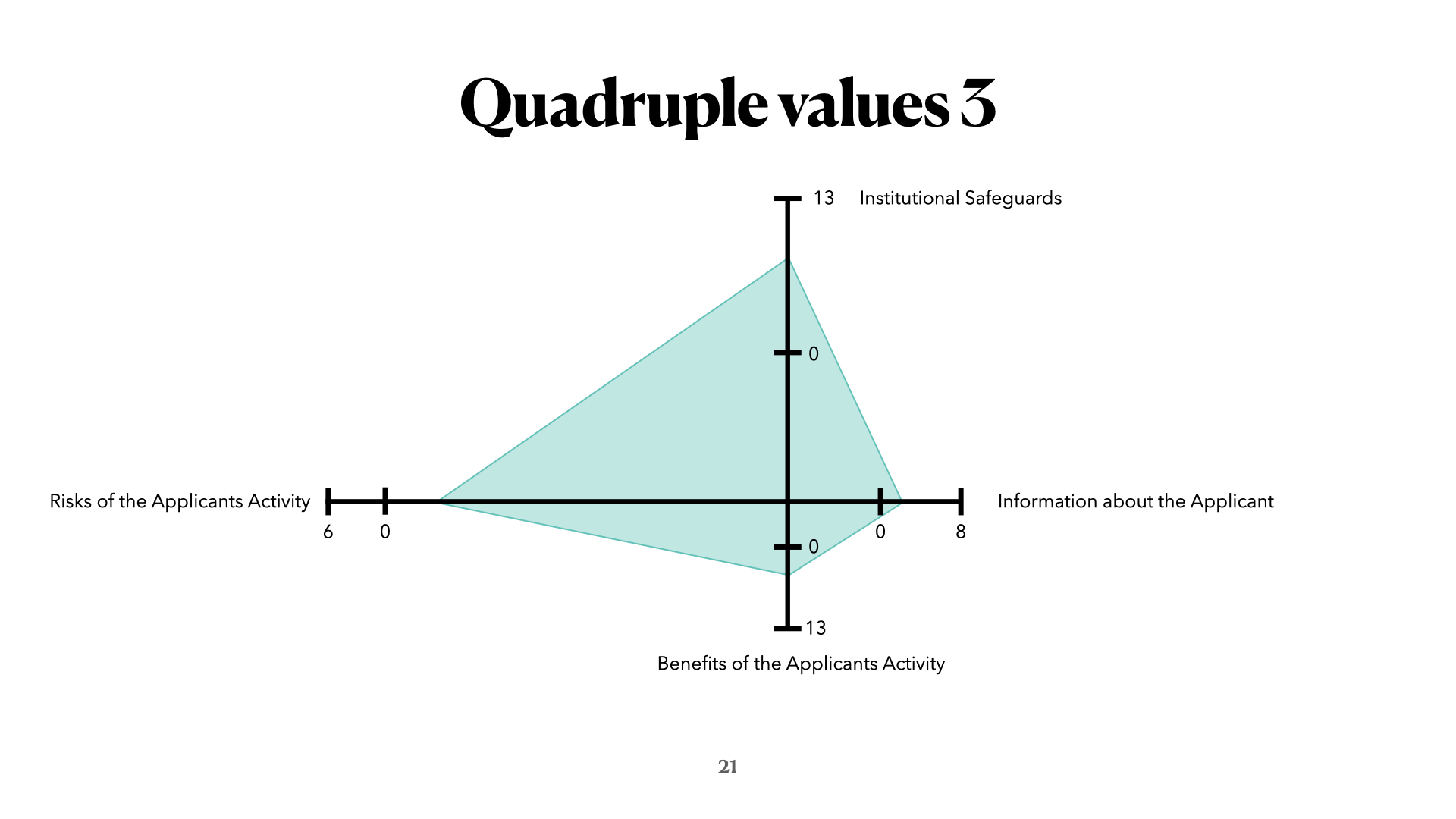}
    \caption{\textbf{Irregular Polygon with Scaled Axes (Quadruple Values 3):} A radar chart with independently scaled axes forms an irregular polygon, emphasizing relative performance in each category. While offering flexibility for different ranges, it may confuse users with its separate axis scales and potential distortion in proportional comparisons. By accommodating varied data ranges, it supports detailed analysis and aligns with project goals for analytical depth. However, additional explanations may be needed to ensure inclusivity and prevent misinterpretation.}
    \label{fig:prototype:quadruplevalues3}
\end{figure}

\begin{figure}[htbp]
    \centering
    \includegraphics[width=0.5\textwidth]{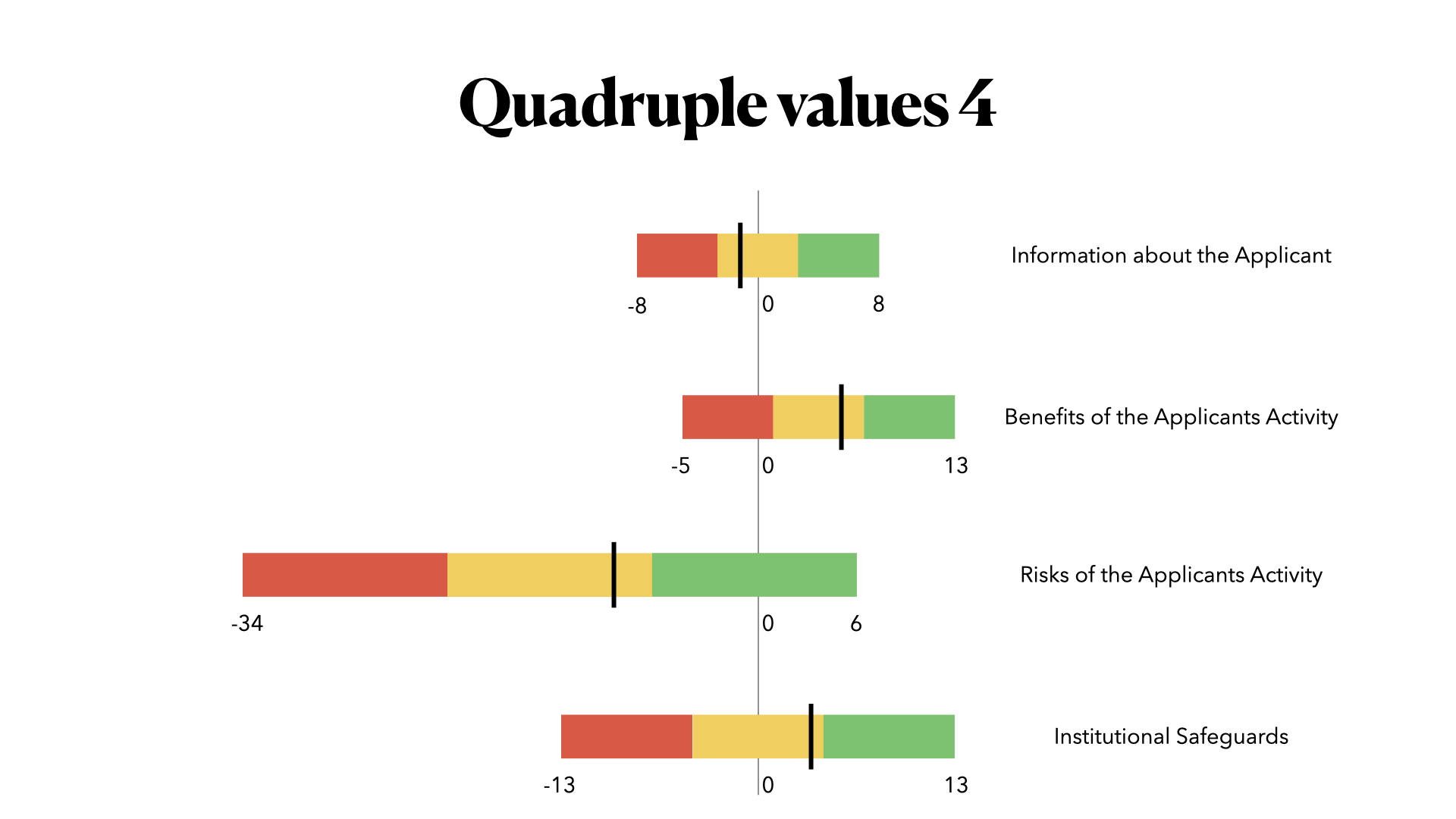}
    \caption{\textbf{Horizontal Bar Chart with Color Coding (Quadruple Values 4):} This chart uses four horizontal bars to represent each dimension, with sections color-coded to indicate positive, neutral, and negative values. It offers a straightforward comparison between dimensions, enhancing interpretability through color coding. While simplifying relationships, it requires colorblind-friendly palettes for accessibility. This design reduces complexity, supports clear categorization, and aids quick interpretation, aligning with project goals and inclusive data experience.}
    \label{fig:prototype:quadruplevalues4}
    
\end{figure}

\begin{figure}[htbp]
    \centering
    \includegraphics[width=0.5\textwidth]{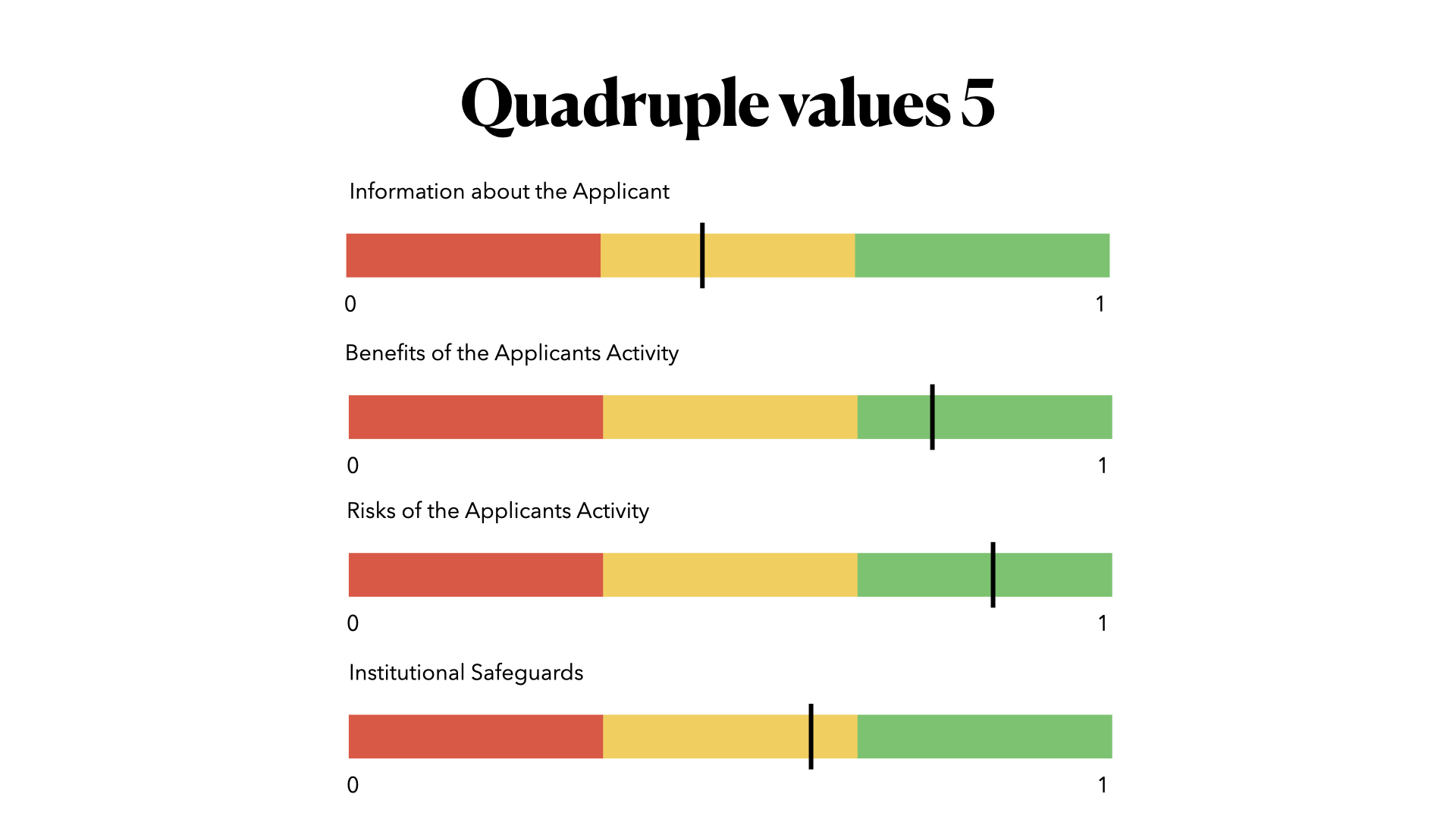}
    \caption{\textbf{Standardized Horizontal Bar Chart (Quadruple Values 5):} Similar to the previous chart, this version normalizes scales (0–1) for consistency across dimensions, simplifying comparisons. The uniform scale facilitates direct comparison but may obscure the magnitude of differences. By maintaining balance and standardization, it aligns with the project's goals for equal data representation and offers an engaging, simple visual format for varied expertise levels.}
    \label{fig:prototype:quadruplevalues5}
\end{figure}

\begin{figure}[htbp]
    \centering
    \includegraphics[width=0.5\textwidth]{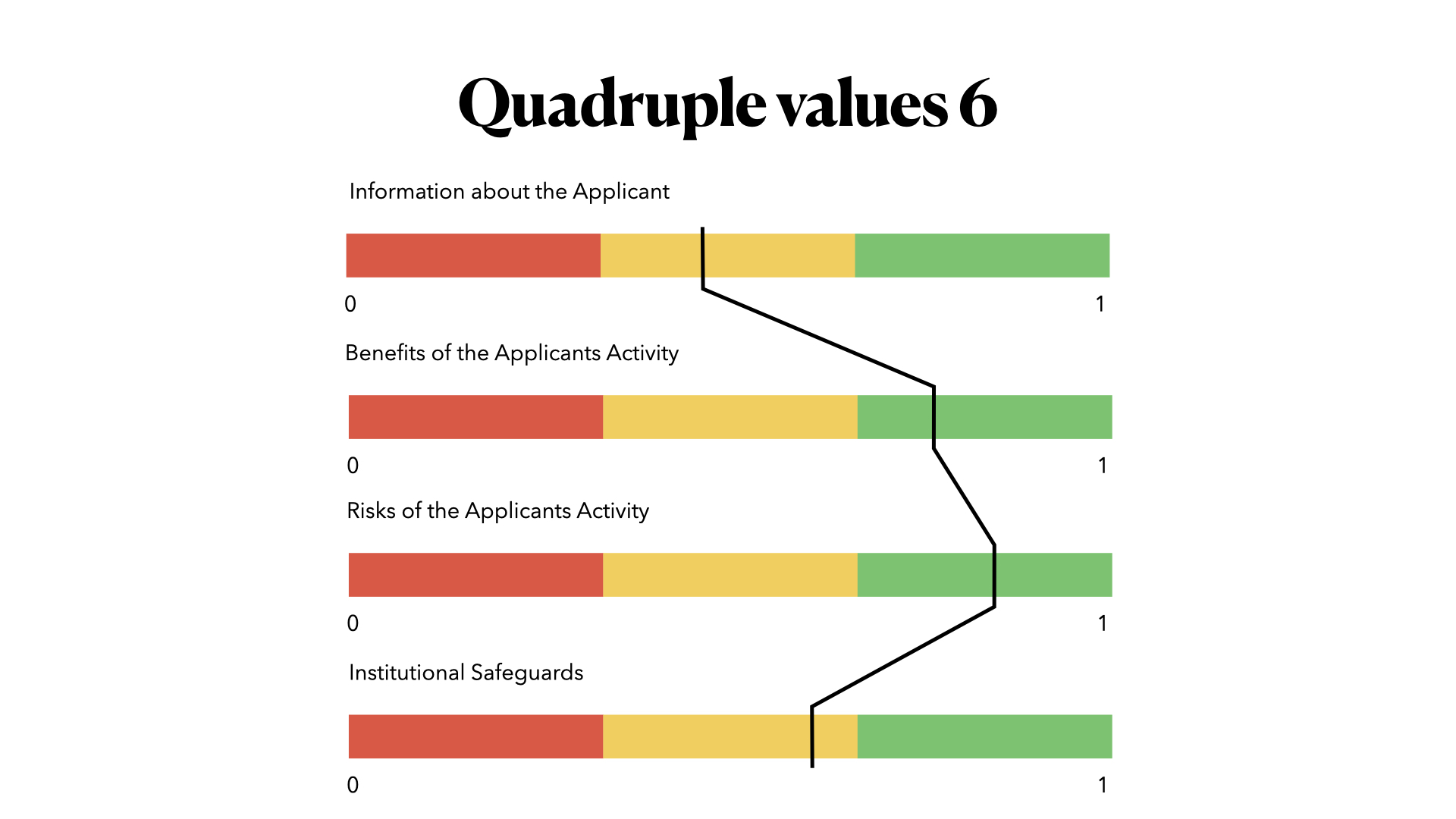}
    \caption{\textbf{Linked Bar Chart (Quadruple Values 6):} This chart features horizontal bars for each dimension, connected by a line to illustrate relationships or flow between values. It highlights patterns and correlations, adding relational insight. While this approach facilitates advanced analysis and aligns with project goals of capturing complex interdependencies, it requires clear annotations to avoid misinterpretation and ensure accessibility, especially for users less familiar with relational visuals.}
    \label{fig:prototype:quadruplevalues6}
\end{figure}

The following design prototypes (\autoref{fig:prototype:singlevalue1} -- \ref{fig:prototype:quadruplevalues6}) are part of the iterative development process of our study. Each prototype represents a distinct approach to simplifying and translating the multi-dimensional aspects of the data into easily interpretable values.

\clearpage

\subsection{Iterative Evaluation}
\label{sec:iterative-evaluation}



\subsubsection{First feedback round}
A sample of purposefully selected stakeholders were sent an email invitation (March 2023) with a private link to the tool. They were asked to try PLUTO (by going through the questions from start to end) and provide comments or suggestions on its practicality, usability, and ethical implications (mostly in writing, with one group meeting us in person). To structure their responses, we provided the following questions: (1)~Was the purpose of the tool clear? (2)~How did you experience the structure of the tool? (3)~Which questions stood out and why? (4)~Were any questions missing, and if so, which ones? One of the primary issues raised at this stage concerned PLUTO's intended target audience. Respondents questioned whether the tool was meant for internal self-assessment within organizations or external assessments, such as audits. In response, we added a single-choice question at the start of the questionnaire, which asks whether the user was assessing their (or their organization's) activities or someone else's, emphasizing that the tool is for individuals and organizations at all levels and can be used for external and self-assessment.  Additionally, respondents recommended streamlining the language of certain questions to accommodate users from diverse knowledge backgrounds, suggesting changes like replacing "applicant" with "data user". In the tool's first iteration, question 8 asked if the data user had a history of ensuring that benefits reached "groups that are granted additional protection by way of anti-discrimination law." Several respondents pointed out that certain populations remain marginalized regardless of legal protections. As one respondent asked:


\begin{quote}{Are [data users] not legally bound by anti-discrimination law anyway? Should the question not look how do they go above and beyond [minimum standards]?}\end{quote}

Based on this feedback, we rephrased the question to ensure it asked the user about benefits to such communities "beyond legal requirements" and, therefore, that it covered groups not explicitly protected by law.  Similarly, question 14 asked if the data use presented risks for "groups afforded special protection under the law (including, but not limited to, children, persons with disabilities, ethnic minorities, sexual minorities)." One respondent suggested that, while compliance with the law is essential, data users may overlook nuanced risks, stating:

\begin{quote}
{… they need to be compliant with the law, but will they have thought the risks through? We fill out whole ethics forms for this. They might assume no risk but we know that things overflow and misfire.}\end{quote}

Respondents also highlighted the need to clarify the distinction between direct and indirect risks to marginalized groups, citing, for example, the immediate harm of a privacy breach versus the indirect risk of biased training data leading to discriminatory algorithms. Question 24 focused on whether a complaints procedure was in place to support those experiencing harm from data use, and if it was "designed in an accessible manner." One respondent suggested making the term "accessible" clearer for the sake of inclusivity, suggesting distinctions such as user-friendly versus disability-friendly access and recommending clearer wording like clear complaints procedure or inclusive access. In response, this question was reworded to specify that "designed in an accessible manner" included complaints procedures that included users with disabilities, whereas "easily accessed" referred to how readily available the procedure is to all users, for example, by being accessible online. Feedback on the tool's visualization aspect was largely positive. One respondent commented:


\begin{quote}{I like the quadrant(s) but I think users of the tool would like to have it spelled out more where they are on the scale (e.g. you are: high risk, low benefit); and customize the feedback and recommendations to them a lot more specifically.}\end{quote}

We decided to include recommendations as part of the PLUTO experience and clarity on the results page regarding the user's final score (e.g., "10 of your answers impact the riskiness rating, and 15 of your answers influence the benefit rating in your result"). Respondents also noted that the method of weighting questions and the assumptions and decisions behind these weightings were unclear. To address this, we included an appendix on the PLUTO page, publicly detailing all the weightings and their rationale. Additionally, in response to queries about specific terms (e.g., public value, risks, benefits, data user), we opted to include a glossary on the PLUTO page, allowing users to refer to it for clarification on the meaning of these terms.

\subsubsection{Second feedback round}

Following the first round of feedback, we sent a follow-up form to the same contacts with an additional question: \emph{Did the results you received match your expectations? Why or why not?} Sixteen of the original 42 respondents took part in the second round of feedback (response rate of 38\%). They noted significant improvements: Most respondents found the tool's purpose clear and reported that it was easier to use. A key technical adjustment was the recommendation to add an information box to each question, offering further clarification to accommodate different users' varying knowledge and expectations and make the questions more widely applicable. For example, one user suggested that the information box could specify what "anti-discrimination law" means, noting that this will vary enormously between countries. Regarding the results page, respondents suggested allowing users to download their results as a PDF. While there was an overall improvement in the clarity of the PLUTO recommendations, some users noted that the wording could be more explicit about what steps they should take to improve their public value score. In response to these comments, we reviewed the wording of the results page and recommendations for clarity and added a function to download the final results. Respondents highlighted that the score on the final results page tended to skew toward Type C (high benefit, high risk). Following a review of the weightings, we concluded that as more questions positively increased the benefits and risk scores, a disproportionate number of users were being directed to Type C, even when answering purposely for a Type A, B, or D result. In response, we decided to normalize the range of results, ensuring a more balanced distribution across the outcome types. This adjustment successfully resolved the issue of skewed results and enabled more accurate reflections of users' input, resulting in a fairer and more representative output.

\clearpage

\subsection{Usability Study Supplements}

\textbf{Data Use Practices of Diagnosys AI}
\label{sec:diagnosys-ai-data-use}

\textbf{Scenario:} Evaluate a proposed project where Diagnosys AI, a private company, uses patient data from the public Capital Region Hospital Network.

\textbf{Activities:} Diagnosys AI will analyze pseudonymized X-ray/CT scans to conduct research improving their AI, aiming to sell a product/service (diagnostic software).

\textbf{Funding \& Reporting:} Diagnosys AI, a medium-sized entity, is funded by income from its own activities (revenue) and a grant from a public body (financing by governmental organisations). They provide an annual public report on project activities (not financials) for the grant. General public access to information relies on GDPR (no legal options to compel disclosure beyond civil or criminal proceedings). They have a procedure in place to respond to requests for information from members of the public (data subjects) seeking access to, or information about the processing of, their personal data.

\textbf{Motivation \& Data:} Uses historical, pseudonymized images. Stated motivations are research (industry-focused) and business and commerce.

\textbf{Impact:} The company argues for benefits covering current and future relevance and that benefits are likely to increase in the future, accruing mainly to Diagnosys AI (commercial) and healthcare provider customers, with potential gains for patients. They use cloud providers with some measures to limit environmental impact.

\textbf{Benefit Distribution:} Focused on high-income countries; no mechanism or intention to benefit low- and middle-income countries (LMICs). Past critiques noted slightly lower AI accuracy for some demographics; they have some, but not a strong or consistent, track record of proactively strengthening benefits for protected groups.

\textbf{Risks:} Foreseeable risks include informational risk to data subjects (the supposedly anonymous data of the patients could potentially be linked back to them), health risk to future patients (from potential AI misdiagnosis), and social risk (AI bias potentially worsening health disparities). Overall likelihood is moderate foreseeable risk for all categories. The data use foreseeably entails elevated risk for protected groups, meaning individuals identified by characteristics like race or gender are more likely to be negatively impacted, potentially due to biases in the data or AI model. There is no information about the specific risks of this project proactively shared with the general public. No specific steps are described to prevent misuse of the algorithms or insights.

\textbf{Safeguards:} Diagnosys AI uses ad-hoc risk assessments internally; findings usually result in information for technical teams. The process involves technical staff and legal experts. There is no dedicated process described for monitoring broader harms post-deployment. Stopping the software use across the network cannot be ended immediately. Patients complain via the hospital; Diagnosys AI has no direct public complaint procedure. Complaints escalated to Diagnosys AI are acknowledged.

\clearpage

\begin{table}[htbp]
\centering
\small
\caption{Probes to guide the results interpretation session.}
\label{tab:probes}
\begin{tabularx}{\columnwidth}{c p{4.2cm} X}
\toprule
\textbf{\#} & \textbf{Moderator Probe} & \textbf{Purpose} \\
\midrule
A1 & \textit{Initial Impression} \newline
First look – what is this page telling you overall? & Assess initial grasp and focus. \\
\addlinespace
A2 & \textit{Visualization Comprehension} \newline
Where does the result show up on the chart? & Test basic chart orientation. \\
\addlinespace
A3 & What do 'Risk' and 'Benefit' on the axes mean here? & Gauge comprehension of the core risk-benefit framework. \\
\addlinespace
A4 & What does the dot's position in this quadrant suggest? & Test understanding of the visual classification system. \\
\addlinespace
A5 & What about these specific numbers for Risk and Benefit? & Check if users link scores to the visual display. \\
\addlinespace
A6 & And this 'Type [X]' classification – what does that imply? & Evaluate comprehension of the data solidarity typology. \\
\addlinespace
A7 & \textit{Input-Output Link} \newline
This text mentions '[X] answers impact risk...' – what does that mean? & Assess if users understand how their inputs generated the output. \\
\addlinespace
A8 & \textit{Recommendation Comprehension} \newline
Let's look at a recommendation... What is this first one suggesting? & Test the clarity and directness of the textual advice. \\
\addlinespace
A9 & \textit{Recommendation Perceived Value} \newline
Is that clear? Does it seem relevant to the scenario? Useful? & Gauge perceived actionability and relevance of advice. \\
\addlinespace
A10 & \textit{Utility Feature Findability} \newline
How would you save or share these results? & Evaluate findability of key utility features. \\
\addlinespace
A11 & \textit{Overall Summary} \newline
In your own words, what's the main outcome for the project according to this? & Assess ability to synthesize information into a conclusion. \\
\bottomrule
\end{tabularx}
\end{table}

\begin{figure}[htbp]
    \centering
    \includegraphics[width=\columnwidth]{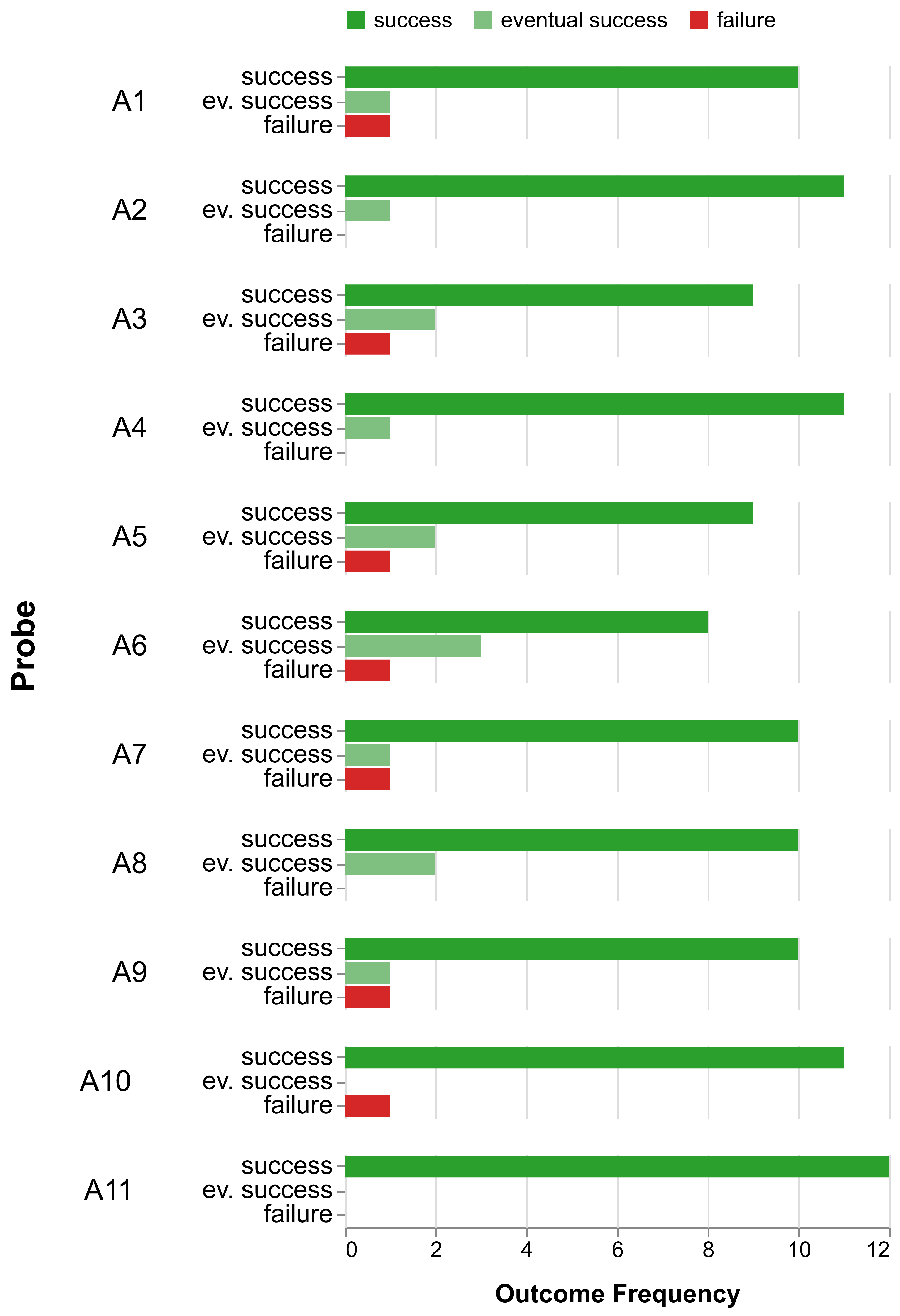}
    \caption{Participant comprehension outcomes ($N=12$) for the 11 structured probes administered during the results interpretation task. Each probe ID (A1-A11) corresponds to the questions detailed in \autoref{tab:probes}. The results show high overall comprehension, with a majority of participants achieving immediate success on most tasks. Notably, all 12 participants successfully synthesized the information to provide an overall summary (A11), indicating the tool's effectiveness in communicating its final assessment.}
    \label{fig:probe_outcomes}
\end{figure}    

\begin{table}[htbp]
\centering
\small
\caption{Subjective usability rating questions.}
\label{tab:ratings}
\begin{tabularx}{\columnwidth}{c p{4.2cm} X}
\toprule
\textbf{\#} & \textbf{Questionnaire Item} & \textbf{Purpose} \\
\midrule
B1 & Overall, how difficult or easy was it to use the PLUTO tool? \newline \textit{(1=Very Difficult, 5=Very Easy)} & Capture perception of effort and interaction fluency. \\
\addlinespace
B2 & How clear or unclear were the questionnaire questions? \newline \textit{(1=Very Unclear, 5=Very Clear)} & Measure comprehensibility of survey content. \\
\addlinespace
B3 & How easy or difficult was it to understand the results visualization? \newline \textit{(1=Very Difficult, 5=Very Easy)} & Assess interpretability of the main visual output. \\
\addlinespace
B4 & How clear or unclear were the textual recommendations? \newline \textit{(1=Very Unclear, 5=Very Clear)} & Evaluate legibility and directness of actionable advice. \\
\addlinespace
B5 & How useful or not useful do you think those recommendations would be? \newline \textit{(1=Not at all Useful, 5=Very Useful)} & Gauge perceived value and applicability of the tool's guidance. \\
\addlinespace
B6 & How confident are you in understanding these results? \newline \textit{(1=Not at all Confident, 5=Very Confident)} & Measure self-assessed certainty in interpretation. \\
\bottomrule
\end{tabularx}
\end{table}

\begin{figure}[htbp]
    \centering
    \includegraphics[width=\columnwidth]{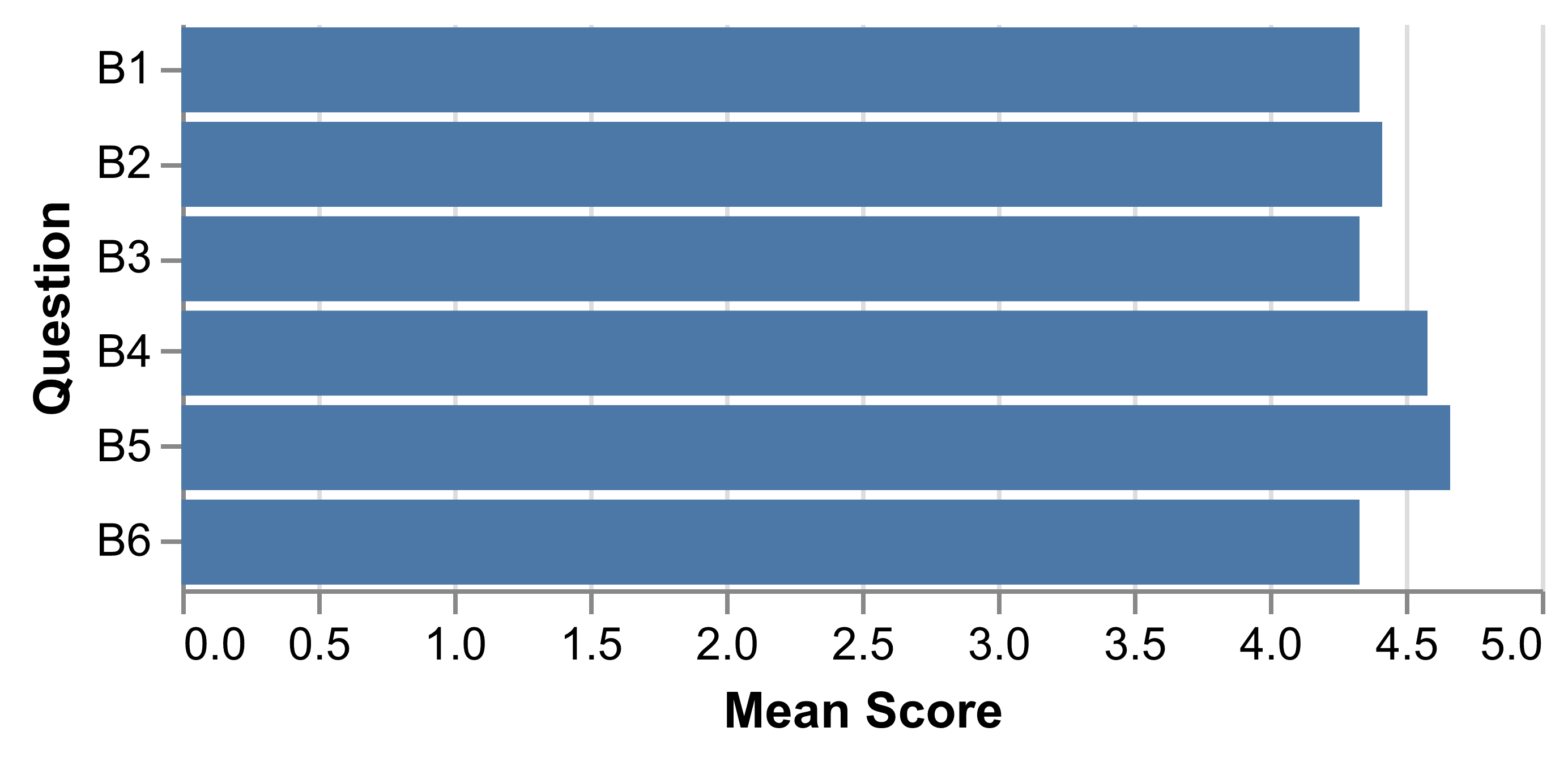}
    \caption{Mean user ratings ($N=12$) on a 5-point Likert scale for key aspects of the PLUTO tool. Each question ID (B1-B6) is detailed in \autoref{tab:ratings}. The consistently high scores across all dimensions provide quantitative support for the tool's usability and effectiveness.}
    \label{fig:quant_results}
\end{figure}

\begin{figure}[htbp]
    \centering
    \includegraphics[width=\columnwidth]{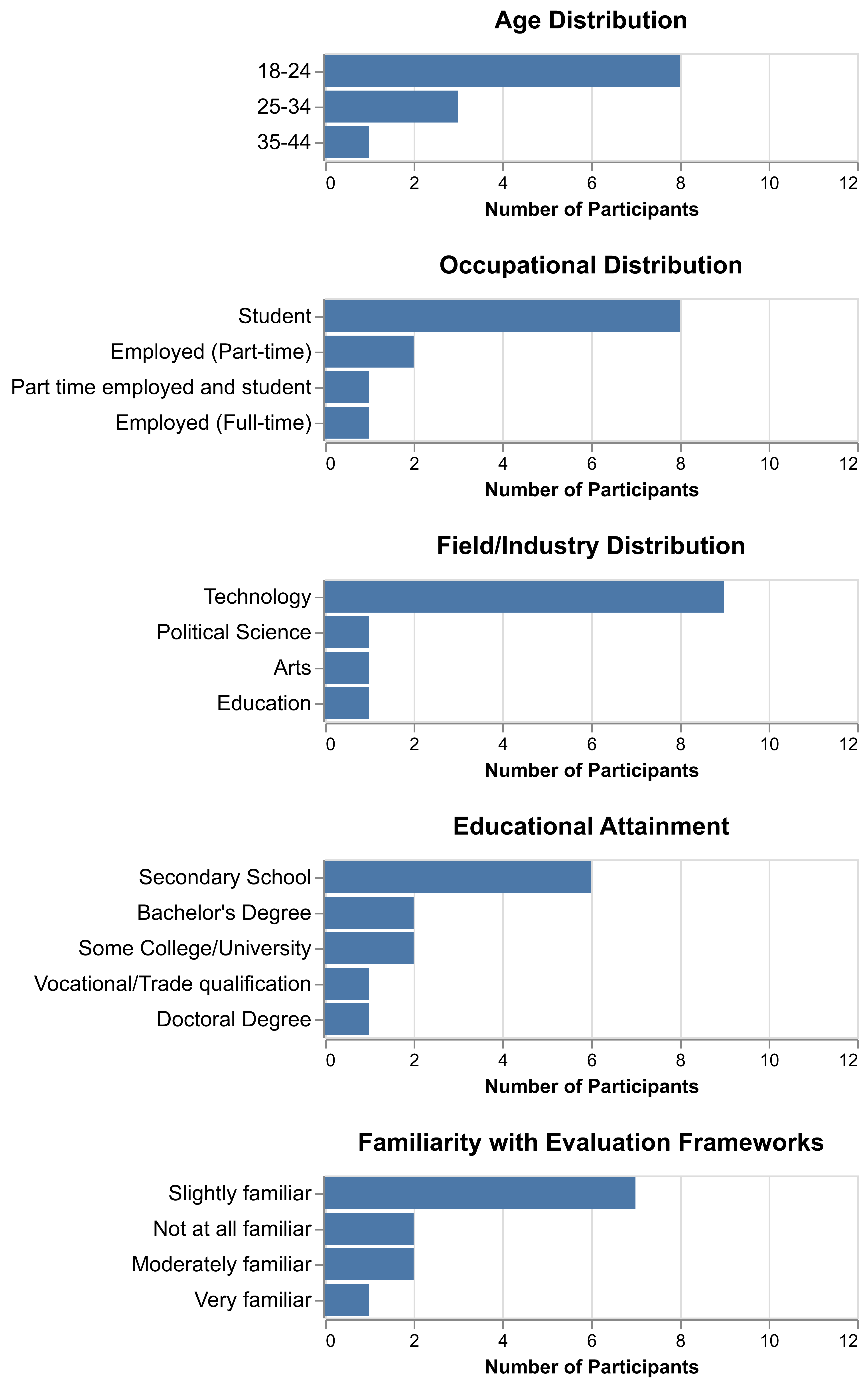}
    \caption{Demographic distribution of the $N=12$ study participants. The data shows participant details across age, occupation, field of work, educational attainment, and prior familiarity with evaluation frameworks. Crucially, the results confirm the cohort was composed of non-specialists, with a majority reporting little to no prior experience in assessing projects for societal impact.}
    \label{fig:demographics}
\end{figure}

\begin{table}[htbp]
\centering
\small
\caption{System Usability Scale (SUS) items, rated on a 5-point Likert scale (1 = Strongly Disagree, 5 = Strongly Agree).}
\label{tab:sus}
\begin{tabularx}{\columnwidth}{c X l}
\toprule
\textbf{\#} & \textbf{Statement} & \textbf{Purpose} \\
\midrule
Q1 & I think that I would like to use this system frequently. & \multirow{10}{1.8cm}[-2.5ex]{To calculate a standard usability score.} \\
Q2 & I found the system unnecessarily complex. & \\
Q3 & I thought the system was easy to use. & \\
Q4 & I think that I would need the support of a technical person to use this system. & \\
Q5 & I found the various functions in this system were well integrated. & \\
Q6 & I thought there was too much inconsistency in this system. & \\
Q7 & I would imagine that most people would learn to use this system very quickly. & \\
Q8 & I found the system very cumbersome to use. & \\
Q9 & I felt very confident using the system. & \\
Q10 & I needed to learn a lot of things before I could get going with this system. & \\
\bottomrule
\end{tabularx}
\end{table}

\begin{figure}[htbp]
    \centering
    \includegraphics[width=\columnwidth]{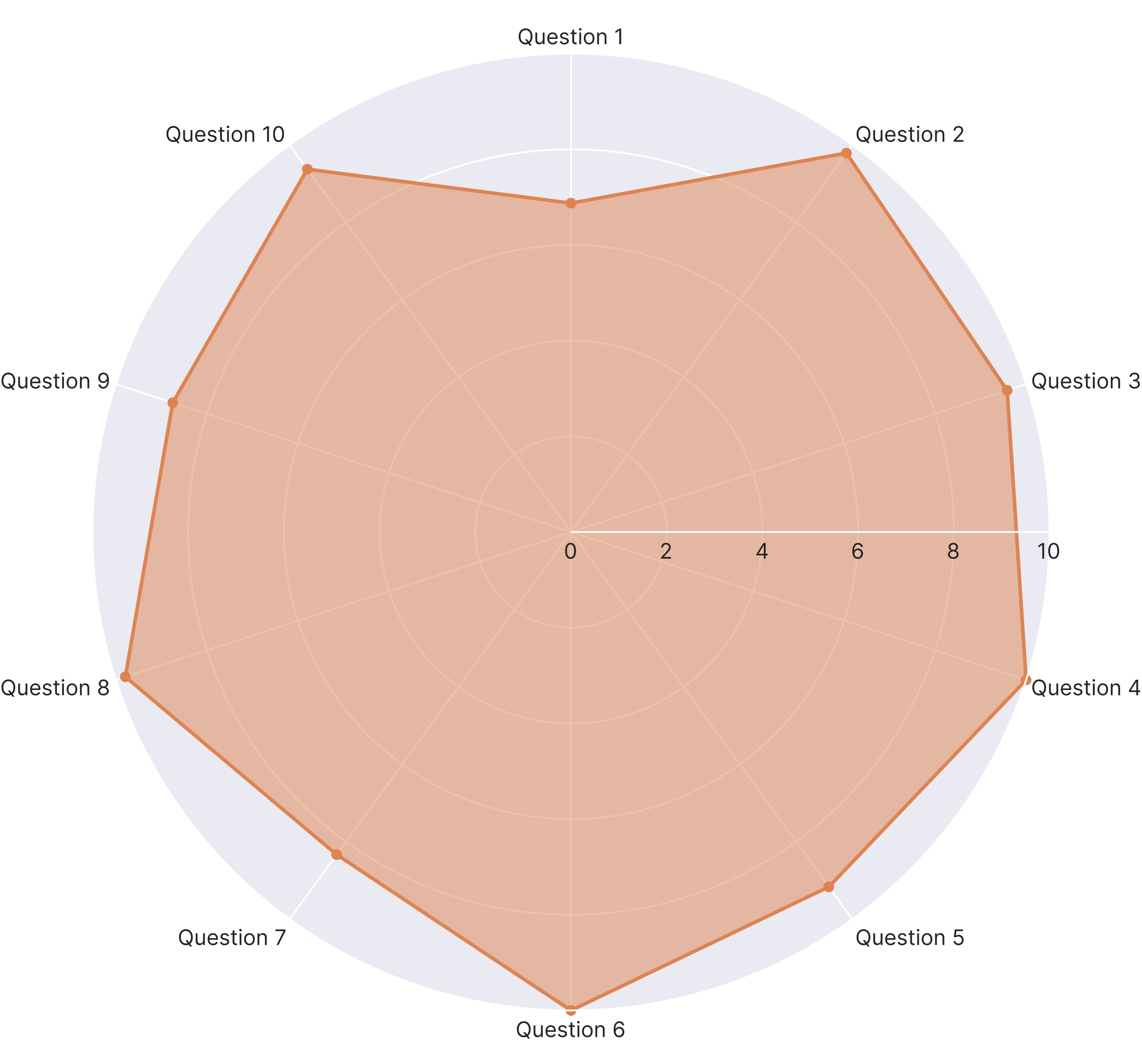}
    \caption{Per-question mean scores for the SUS questionnaire (see \autoref{tab:sus} for question text). The strong positive ratings for items related to ease of use (Q3), confidence (Q9), and learnability (Q7) align with the high overall score. The lower rating for Q1 (``I think that I would like to use this system frequently'') was consistently explained by participants as a reflection of the task's specific, infrequent nature, not as a critique of the tool itself.}
    \label{fig:sus_per_question}
\end{figure}

\begin{figure}[htbp]
    \centering
    \includegraphics[width=\columnwidth]{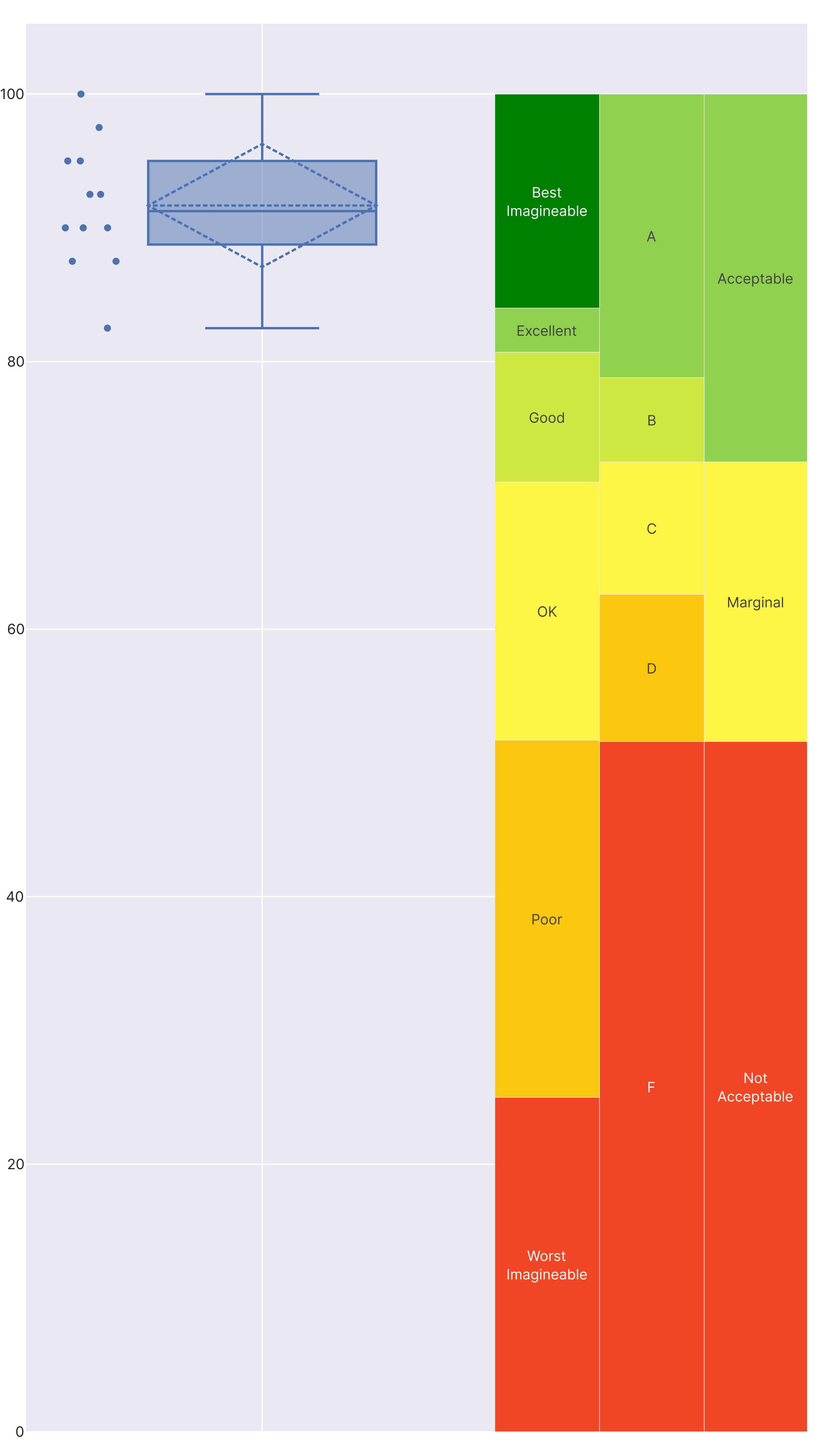}
    \caption{Distribution of System Usability Scale (SUS) scores from the 12 participants. The mean score of 91.67 places the tool in the ``Best imaginable'' range for perceived usability, with a tight distribution indicating consistently positive experiences.}
    \label{fig:sus_results}
\end{figure}

\begin{table*}[htbp]
\centering
\small
\caption{Probes for the concluding semi-structured interview.}
\label{tab:interview}
\begin{tabularx}{\textwidth}{c X X}
\toprule
\textbf{\#} & \textbf{Moderator Probe} & \textbf{Purpose} \\
\midrule
C1 & What was your overall impression of the PLUTO tool? & Elicit a summary judgment and top-of-mind thoughts. \\
\addlinespace
C2 & What was particularly easy? What was most difficult? & Identify specific strengths and weaknesses in the UI and flow. \\
\addlinespace
C3 & Did the questionnaire options fit the scenario well? & Assess perceived completeness and applicability of content. \\
\addlinespace
C4 & Based on using it, what is the main purpose of PLUTO? & Check if the tool's core value proposition is communicated. \\
\addlinespace
C5 & Who do you imagine using this tool? Could a non-expert use it? & Gauge perceived inclusivity and accessibility for the target audience. \\
\addlinespace
C6 & Do you have any final comments or suggestions for improving PLUTO? & Collect user-driven ideas for future design iterations. \\
\bottomrule
\end{tabularx}
\end{table*}

\begin{table*}[htbp]
\centering
\small
\caption{Selected thematic quotes from the usability study, illustrating the core user experiences.}
\label{tab:thematic_quotes}
\begin{tabularx}{\textwidth}{p{2.5cm} p{7.5cm} X}
\toprule
\textbf{Theme} & \textbf{Quote} & \textbf{Relevance} \\
\midrule
\textbf{\parbox{2.5cm}{Agency \& \\ Empowerment}} & \textit{``My takeaway as a patient is that I would not give them my data because it's discouraged... Personally, the outcome for me would be to not give them my data.''} & Shows the user making a definitive, personal decision based on the tool's assessment, demonstrating direct empowerment. \\
\addlinespace
& \textit{``If this were my company, I would have to think heavily about better carving out and communicating the future benefits of this company.''} & The user adopts a decision-maker's perspective, considering concrete actions for project improvement based on the tool's results. \\
\addlinespace
& \textit{``If I was evaluating it from a government perspective, I would probably look at these recommendations and maybe try to implement some thresholds...''} & Illustrates empowerment at a policy level, where the user envisions using the tool's output to inform regulation. \\
\midrule
\textbf{\parbox{2.5cm}{Explainability \& \\ Trust}} & \textit{``The tool affirmed what I initially had in mind... So I feel pretty confident that the tool aligned with what I initially had in mind.''} & Shows trust being built when the tool's logic validates the user's own intuition about the scenario. \\
\addlinespace
& \textit{``That reminds me of the question where I was like, 'Oh, research for industry or for science..?'... if I had clicked science, that probably would have been counted as a bigger benefit.''} & Demonstrates the user understood the causal link between their specific input and the final score, a core aspect of explainability. \\
\addlinespace
& \textit{``I would expect the company that wants my data to give me this information and proof... I would expect a filled out Pluto survey from them.''} & The user reframes the tool as a mechanism for corporate transparency, showing trust in its framework to hold others accountable. \\
\addlinespace
& \textit{``I would kind of like to know how the recommendations are generated.''} & User curiosity about the tool's logic indicates trust in the output. \\
\midrule
\textbf{\parbox{2.5cm}{Learning \& \\ Reflection}} & \textit{``I was somewhat surprised... I still perceived the value to be quite high... But I guess the way the tool works, it classified it to be lower because of the harm... When I think about it now, it makes sense to me.''} & Captures a moment of reflection where the user re-evaluates an initial bias after the tool highlights a previously overlooked factor. \\
\addlinespace
& \textit{``They haven't really, really thought about how they use the data in depth... It seems like they are thinking a lot about their technology; that's what I understand from the results.''} & Illustrates a key takeaway where the user infers the project's priorities (technology over societal impact) from the assessment results. \\
\addlinespace
& \textit{``Although the risk... isn't that high and the benefit... isn't that low, the combination... makes it discouraging. It has too many risks for the quite low benefit.''} & The user articulates the core risk-benefit trade-off, demonstrating a grasp of the tool's central evaluative concept. \\
\midrule
\textbf{\parbox{2.5cm}{Usability \& \\ Interpretation}} & \textit{``I would have expected the edit button to either take me directly to the question... it made me feel like, 'Oh, well, that's just kind of pointless.'''} & Provides specific, actionable feedback on a UI feature whose behavior violated user expectations, causing friction. \\
\midrule
\textbf{Visualization Comprehension} & \textit{``The best would be having high benefit and low risk, which would be somewhere in this top-left corner. And the worst would be if we have a use case where there is high risk and low benefit.''} & The user correctly interprets the 2D risk-benefit matrix, including the ideal and worst-case quadrants. \\
\addlinespace
& \textit{``It could initially be a little bit confusing that low risk is good and high risk is bad... But the coloring makes it very intuitive for me.''} & Identifies a potential confusion with axis orientation but confirms that a specific design choice (color-coding) effectively resolved it. \\
\bottomrule
\end{tabularx}
\end{table*}

\end{document}